\documentclass[12pt]{article}

\usepackage{rotating}

\usepackage{multirow}
\usepackage{amsmath,amssymb,graphicx, graphics, color,verbatim} %use drftcite in draftmode
\usepackage{cite}
\usepackage{latexsym} 
\usepackage{cancel}
\usepackage{hyperref}

\usepackage{enumerate} 

\usepackage[normalem]{ulem} %for strikethrough

\pdfoutput=1
\usepackage{epsf}
\usepackage{pstricks} 
\usepackage{comment}

\usepackage{array}

\newcommand{\centeron}[2]{{\setbox0=\hbox{#1}\setbox1=\hbox{#2}\ifdim
\wd1>\wd0\kern.5\wd1\kern-.5\wd0\fi \copy0
\kern-.5\wd0\kern-.5\wd1\copy1\ifdim\wd0>\wd1
                                   \kern.5\wd0\kern-.5\wd1\fi}}
\newcommand{\ltap}{\>\centeron{\raise.35ex\hbox{$<$}}
                           {\lower.65ex\hbox{$\sim$}}\>}
\newcommand{\gtap}{\>\centeron{\raise.35ex\hbox{$>$}}
                           {\lower.65ex\hbox{$\sim$}}\>}
\newcommand{\gsim}{\mathrel{\gtap}}
\newcommand{\lsim}{\mathrel{\ltap}}

\newcommand\ZZ{\hbox{\zfont Z\kern-.4emZ}}
\font\zfont = cmss10 %scaled \magstep1

\newcommand{\fref}[1]{Fig.\ \ref{f.#1}}
\newcommand{\eref}[1]{Eq.\ (\ref{e.#1})}

\newcommand{\sref}[1]{Section \ref{s.#1}}
\newcommand{\ssref}[1]{Section \ref{ss.#1}}

\newcommand{\cref}[1]{Chapter \ref{c.#1}}

\newcommand{\ba}{\begin{array}}
\newcommand{\ea}{\end{array}}

\newcommand{\beq}{\begin{eqnarray}}% can be used as {equation} or  {eqnarray}
\newcommand{\eeq}{\end{eqnarray}}
\newcommand{\beqs}{\begin{eqnarray*}}
\newcommand{\eeqs}{\end{eqnarray*}}

\newcommand{\bal}{\begin{align}} %this is glorious: arbitrary number of columns, alignment
\newcommand{\eal}{\end{align}}

\def\bi{\begin{itemize}}
\def\ei{\end{itemize}}
\def\ben{\begin{enumerate}}
\def\een{\end{enumerate}}
\def\bc{\begin{center}}
\def\ec{\end{center}}
\def\bt{\begin{table}}
\def\et{\end{table}}
\def\btb{\begin{tabular}}
\def\etb{\end{tabular}}

\def\gev{\, {\rm GeV}}

\def\tev{\, {\rm TeV}}

\def\mass2{mass${}^2$}

\def\simgt{\stackrel{>}{{}_\sim}}

  \newcommand{\ww}{
$W^+W^-$
}

\newcommand{\ifb}{\mathrm{fb}^{-1}}

\newcommand{\met}{\mathrm{MET}}

\newcommand{\hgg}{$h\rightarrow \gamma\gamma$~}

\begin{document}
\bibliographystyle{unsrt}
\begin{titlepage}
\begin{flushright}
\small{YITP-SB-13-12}
\end{flushright}

\vskip2.5cm
\begin{center}
\vspace*{5mm}
{\huge \bf Casting Light on BSM Physics with\vspace{3mm} \\  SM Standard Candles}
\end{center}
\vskip0.2cm

\begin{center}
{\bf David Curtin$^{1}$, Prerit Jaiswal$^{2}$, Patrick Meade$^{1}$, Pin-Ju Tien$^{1}$}

\end{center}
\vskip 8pt

\begin{center}
{\it $^1$ C. N. Yang Institute for Theoretical Physics\\ Stony Brook University, Stony Brook, NY 11794.}\\
\vspace*{0.3cm}
{\it $^2$ Department of Physics, Florida State University, Tallahassee, FL 32306 }\\
\vspace*{0.3cm}

\vspace*{0.1cm}

{\tt curtin@insti.physics.sunysb.edu, prerit.jaiswal@hep.fsu.edu, meade@insti.physics.sunysb.edu, tien@insti.physics.sunysb.edu}
\end{center}

\vglue 0.3truecm

\begin{abstract}

The Standard Model (SM) has had resounding success in describing almost every measurement performed by the ATLAS and CMS experiments. 
In particular, these experiments have put many beyond the SM models of natural Electroweak Symmetry Breaking into tension with the data. 
It is therefore remarkable that it is still the LEP experiment, and not the LHC, which often sets the gold standard for understanding  the possibility of new color-neutral states at the electroweak (EW) scale.
Recently, ATLAS and CMS have started to push beyond LEP in bounding heavy new EW states, but a gap between the exclusions of LEP and the LHC typically remains.
In this paper we show that measurements of SM Standard Candles can be repurposed to set entirely complementary constraints on new physics.
To demonstrate this, we use \ww cross section measurements to set bounds on a set of slepton-based simplified models which fill in the gaps left by LEP and dedicated LHC searches. 
Having demonstrated the sensitivity of the \ww measurement to light sleptons, we also find regions where sleptons can improve the fit of the data compared to the NLO SM \ww prediction alone. Remarkably, in those regions the sleptons also provide for the right relic-density of Bino-like Dark Matter and provide an explanation for the longstanding $3\sigma$ discrepancy in the measurement of $(g-2)_\mu$.
\end{abstract}

\end{titlepage}

%%%%%%%%%%%%%%%%%%%%%%%%%%%%%%%%%%%%%%%%%%%%%%%%%%%%%%%%%%%%%%%%%%%%%%
%%%%%%%%%%%%%%%%%%%%%%%%%%%%%%%%%%%%%%%%%%%%%%%%%%%%%%%%%%%%%%%%%%%%%%% 
\section{Introduction}
\label{s.intro} \setcounter{equation}{0} \setcounter{footnote}{0}
%%%%%%%%%%%%%%%%%%%%%%%%%%%%%%%%%%%%%%%%%%%%%%%%%%%%%%%%%%%%%%%%%%%%%%
%%%%%%%%%%%%%%%%%%%%%%%%%%%%%%%%%%%%%%%%%%%%%%%%%%%%%%%%%%%%%%%%%%%%%

The LHC has made amazing strides forward in understanding the SM at the TeV scale, and what could possibly lay beyond the SM.  However, the LHC as with all hadron colliders, produces colored states with much larger cross sections than Electroweak (EW) states of the same mass.   The larger colored cross sections typically imply that the reach of new physics is much higher for colored states, while new EW states have much weaker bounds and discovery prospects.  For more than a decade the results of the LEP experiments have generically set the most stringent constraints on new EW states beyond the SM despite the full run of the Tevatron physics program.  However, with the large energy and luminosity of the LHC, ATLAS and CMS have started to probe and set constraints beyond LEP for very generic EW final states~\cite{ATLASewbounds, cmssleptons , cms8ew3lep, atlas8ew3lep}.

Despite the impressive achievements of the LHC, we are still relatively insensitive to new EW physics with mass $\mathcal{O}(100)\gev$.  At this energy range, backgrounds for new physics are dominated by SM gauge boson processes, and thus kinematical handles that searches rely on to separate signal from background are much less powerful.  This typically results in a gap in searches/exclusions between LEP~\cite{LEPlimits, pdg:2012} and the new ATLAS and CMS bounds~\cite{ATLASewbounds, cmssleptons, cms8ew3lep, atlas8ew3lep} which we will further illustrate in Section~\ref{s.bounds}.  This gap will not be filled simply with increased luminosity or energy, but will require more effort focused on understanding SM standard candles such as diboson production.  In this paper we investigate the possibility of using measurements of these standard candles to directly bound new physics.  This has been done by the experiments in looking for anomalous triple gauge boson couplings, but these deviations typically appear in tails of distributions and are rather straightforward to search for. 
We will show in \sref{bounds} how the gap between the LEP and LHC slepton searches can {\em already} be filled in with existing low luminosity LHC run I measurements of the SM (up to a very interesting region we discuss later).  Following this program of investigating SM standard candles is clearly one of paramount importance if we want to fully understand physics associated with the EW scale.  In turn, this means that ATLAS and CMS should make just as much of an effort to measure as many fully differential SM EW processes as they do for searches.  It is also crucial for the experiments to be cognizant of the fact that new physics is not relegated only to exotics, SUSY searches or extreme kinematics, and can contaminate SM measurements or even improve them in some cases~\cite{Curtin:2012nn,Lisanti:2011cj}.

An interesting example of the importance of SM standard candles from this past year is the measurement of the SM \ww cross section.  At 7 TeV both ATLAS and CMS measured this cross section to be higher than the SM result \cite{atlas7ww, cms7ww}, but very consistent with each other. At 8 TeV using $3.5\ifb$ of data CMS \cite{cms8ww} measured this cross section to be even more discrepant with the SM than at 7 TeV.   Combining these measurements in the most straightforward manner yields a $3\sigma$ excess.  This of course is not a smoking gun for new physics, but it was shown that new physics can  significantly improve the $\chi^2$ for all differential measurements of these cross sections~\cite{Curtin:2012nn}.    The simplest new physics explanation of the \ww cross section increase is to introduce a new source of $W$ gauge bosons. The challenge lies in avoiding additional particle production, since the jet veto and strict OS dilepton cut used in the \ww measurements basically requires the new physics to produce nothing but $l^+l^-+\met$ in the final state. 
This avenue was explored in \cite{Curtin:2012nn}, where Chargino pair production followed by the subsequent decay to $W$ gauge bosons and $\met$ was studied \footnote{There is a small contribution to same-sign dilepton production from $\chi_1^+ \chi_1^0$ channel. Such a signature is absent if the neutralino is Dirac, as pointed out in \cite{DiracEWkino}}.      Another direction based on stop production was explored in~ \cite{Rolbiecki:2013fia}, where it was demonstrated that the process $pp\rightarrow \tilde{t}\tilde{t}^*\rightarrow b\bar{b}\chi^+\chi^-\rightarrow b\bar{b}W^+W^-\chi^0\chi^0$ can contribute to the \ww cross section provided that the $b$ jets are soft enough.  

In this paper we explore an alternate scenario for explaining the measured \ww cross section without producing actual $W$ gauge bosons. In a SUSY model where sleptons are light enough to be directly produced at the LHC, their subsequent decay $\tilde{l}\rightarrow l \chi^0$ can also contribute to the $l^+l^-+\met$ final state.  Typically this spectrum will be harder than in scenarios with $W$ partners as previously described since the $\met$ will come from 2 missing particles rather than being spread amongst 4.  However, as we will show there is a region of $m_{\tilde{l}}-m_{\chi^0}$ parameter space where similar to \cite{Curtin:2012nn,Rolbiecki:2013fia} the $\chi^2$ of the SM \ww cross section can be significantly improved.  There is an important quantitative and qualitative difference between slepton and chargino based models.  In slepton models only the flavor diagonal contribution is realized in the \ww cross section measurement, whereas charginos will contribute exactly as SM $W$'s with respect to flavor.  At this point the (publicly available) data from the \ww cross section in the flavor separated measurements is not sufficient to favor either of these two possibilities over one another, but it will be an important phenomenological handle in the future.  

Light sleptons can also cast light on several other important puzzles. 
In models with neutralino dark matter (DM) that is mostly Bino, light sleptons are required to achieve the correct relic density (unless the Bino is tuned to lie close to the Higgs resonance region or the sneutrino co-annihilation region). They are also favored as an explanation to the anomalous measurement of $(g-2)_\mu$ \cite{pdg:2012}.  Finally, if the \hgg rate is higher than the SM rate as hinted at by the ATLAS measurement \cite{ATLAS:Higgsdiphoton}, then light $\tilde{\tau}$'s are a possible explanation.  
We will show in this paper that, remarkably, the same region of slepton-bino parameter space preferred by the \ww cross section measurement also naturally accounts for the correct dark matter relic density  $(g-2)_\mu$. A slight increase in the \hgg rate can also be accommodated at the price of introducing violations of Lepton Flavor Universality (LFU) at a level well below experimental constraints on LFU violation, and dark matter direct detection cross sections are below current bounds set by XENON100 \cite{xenon100} but would be discovered by the next generation of experiments.

The structure of the paper is organized as follows. In Section 2, we use the \ww SM standard candle as an example to shed light on a simplified model of direct slepton production that is particularly difficult to detect at the LHC.  We show that the gap that currently exists between the LEP and LHC studies can effectively be snuffed out over a large range of simplified model parameter space.  We also show that in one particular simplified model, LEP {\em still} sets the world's best bound.  In certain regions of the parameter space, we show that the simplified model can actually improve the \ww cross section measurements agreement between data and theory using the aforementioned slepton model.  In Section 3, we explore the ramifications of this particular part of parameter space where new physics can improve the fit to data, including DM, $(g-2)_\mu$, and \hgg and show that their separate preferred regions to explain all these disparate phenomena is actually the same region in model space.  Finally, we discuss the possibility for future uses of SM standard candles and the general program of investigating the only scale for EW physics at which we have found new particles thus far.

%%%%%%%%%%%%%%%%%%%%%%%%%%%%%%%%%%%%%%%%%%%%%%%%%%%%%%%%%%%%%%%%%%%%%%
%%%%%%%%%%%%%%%%%%%%%%%%%%%%%%%%%%%%%%%%%%%%%%%%%%%%%%%%%%%%%%%%%%%%%%%
\section{New Electroweak Bounds from \ww Cross Section}
\label{s.bounds} \setcounter{equation}{0} \setcounter{footnote}{0}
%%%%%%%%%%%%%%%%%%%%%%%%%%%%%%%%%%%%%%%%%%%%%%%%%%%%%%%%%%%%%%%%%%%%%%
%%%%%%%%%%%%%%%%%%%%%%%%%%%%%%%%%%%%%%%%%%%%%%%%%%%%%%%%%%%%%%%%%%%%%

 % summarizing the WW measurement 
Both ATLAS and CMS measure the \ww cross section by counting the number of events in the $l^+l^-+\met$ final state, intending to capture mostly \ww$\rightarrow l^+\nu\l^-\bar \nu$ decays.  There are other SM contributions to this final state: the contribution of $t \bar t$ and other QCD final states is reduced with a jet veto, while $m_{ll}$ and $p_T$ cuts are used to reduce the Drell-Yan contribution and further isolate the \ww contribution. As has been demonstrated in \cite{ Curtin:2012nn,Rolbiecki:2013fia}, even with these cuts it is possible for BSM events to contaminate the $l^+l^-+\met$ final state. This leads to significant deviations in both the measured overall \ww cross section as well as the shape of the associated kinematic distributions. In this section we will explore the utility of the \ww measurement not as a Standard Model Standard Candle, but as a \emph{search} for BSM physics. This has only become feasible at the LHC, both due to the high statistics of the measurement as well as the low theoretical errors on the modern \ww cross section prediction, which are now interpreted as a SM background.

Sleptons have low production cross sections and are difficult to study at the LHC. That makes them a natural test bed for our methods. 
The typical mass scale for slepton bounds prior to the LHC was set by LEP-II at approximately $100\gev$~\cite{LEPlimits}, with some variation depending on the particular flavor of slepton. 
Most of the early LHC bounds on sleptons were based on producing them in cascade decays from new EW states.  These bounds typically constrained heavier mass sleptons, but depended crucially on other parts of the BSM spectrum. 
Recently, however, CMS has set a bound on \emph{direct} LH slepton production~\cite{cmssleptons} which  complement LEP in a different region of the neutralino-slepton mass plane (assuming degenerate LH selectrons and smuons). 
This is done by exploring the difference in kinematics between SM \ww and sleptons using the $M_{CT\perp}$ variable, which essentially encodes the mass scale separation of the mother particles that produce the charged leptons and the particles that make up the $\met$ in the event.  In the case of direct slepton production and decay to $l^\pm \tilde \chi_1^0$, if there is a larger mass splitting than the background which is dominated by SM $W$'s, then strong bounds can be set on the slepton mass.  
However, in the region where the slepton-neutralino mass scale separation becomes more similar to $m_W$, the bounds disappear because of the large SM backgrounds.   

In summary, slepton bounds from LEP are relatively insensitive to $m_{\tilde \chi^0_1}$ but only go as high as 100 GeV, while slepton searches at the LHC probe higher masses but loose sensitivity for $m_{\tilde l} - m_{\tilde \chi^0_1} \lsim m_W$.  However, it is exactly in this ``$WW$-like funnel'' that the \ww cross section measurement would be most sensitive to contamination by slepton decay products. This motivates using the cross section measurement to set bounds in order cover the entire slepton-neutralino mass plane.

\fref{slepbounds} shows our results. We derived slepton mass bounds for selectrons and smuons at the same mass decaying into a neutralino, for each of the performed LHC \ww measurements and all of them combined. To do this we create a grid of SLHA spectrum files with decay tables using \texttt{CPsuperH 2.3} \cite{cpsuperh} and simulate direct slepton production\footnote{Our \texttt{FastJet 3.0.2}  \cite{fastjet} based analysis code took into account lepton isolation requirements and geometrical acceptances but did not simulate detector effects. Given the nature of our final state this will not invalidate our results.} using \texttt{Pythia 6.4/8.15}~\cite{pythia} (hard process/shower), normalized to the NLO cross section calculated using \texttt{Prospino 2.1}~\cite{prospino}.
Since the source of the observed \ww excess is unclear, we obtained slepton limits in two ways. The solid lines indicate bounds obtained by analyzing only the \emph{shape} of the various kinematic distributions in the \ww measurements, normalizing the SM theoretical expectation so that the SM + BSM overall expected event count matched the measurement. This bound should be very robust, even in light of possible future corrections to the SM \ww cross section calculation (unless they significantly change the expected shape of kinematic distributions). On the other had, the dashed lines show bounds obtained by comparing kinematic distributions of expected SM + BSM to the data, without renormalizing the SM prediction. These bounds can be significantly more powerful, but might be less robust. Comparing the two bounds can be instructive. 

We show those bounds along-side LEP and CMS bounds\footnote{The explicit cross section bounds in \cite{cmssleptons} are not of high enough resolution in the slepton-neutralino mass plane to compute a useful mass exclusion curve for the LH + RH slepton case. (It is enough to show, however, that CMS sets no bounds on RH sleptons only.) Therefore we use the CMS supplied LH slepton mass bound in the LH + RH plot as well, which makes it slightly conservative: the magenta region should be roughly $\mathcal{O}(5 \gev)$ `larger' in every direction than shown.}, and the complementarity of our $WW$-derived bounds is clear -- they help fill in the $WW$-like funnel, inaccessible to both LEP and CMS dedicated slepton searches. We can see that the gap between the LEP measurement and the LHC can now start to be closed. Our bounds also represent the first LHC bounds on direct RH-only slepton production for degenerate selectrons and smuons, but the LHC still cannot set any bounds for just a single RH slepton generation.

For the 7 TeV ATLAS and CMS measurements, the bounds obtained by our two analysis methods are almost identical. This is due to the relatively small size of the $WW$-excess in those measurements, compared to the slepton  cross section dependence on slepton mass. The situation is very different for the 8 TeV CMS measurement: while the shape-only bound looks as one might expect, the bound obtained without renormalizing the SM expectation seems to `exclude' the entire slepton mass plane with the exception of a small island centered around $m_{\tilde \ell} \sim 100 \gev, m_{\tilde \chi^0_1} \sim 60 \gev$. This is not due to any extraordinary exclusion power of the CMS8 measurement, but rather because the measurement of the \ww cross section is so high that, under the assumptions of the un-renormalized analysis, the Standard Model itself is excluded at better than 95\% CL. Only within the small island is the slepton contribution so large as to push the $p$-value of the kinematic fit above 0.05. 

This result underscores the utility of setting bounds conservatively, using the shape of kinematic distributions only, but also serves as motivation for going one step further: using the \ww measurement not just for exclusion, but for \emph{discovery} of a possible BSM signal.

\begin{figure}[htbp] %  figure placement: here, top, bottom, or page
\vspace*{-1.5cm}
\begin{center}
\hspace*{-1.5cm}
 \begin{tabular}{cc}
 \includegraphics[width=9cm]{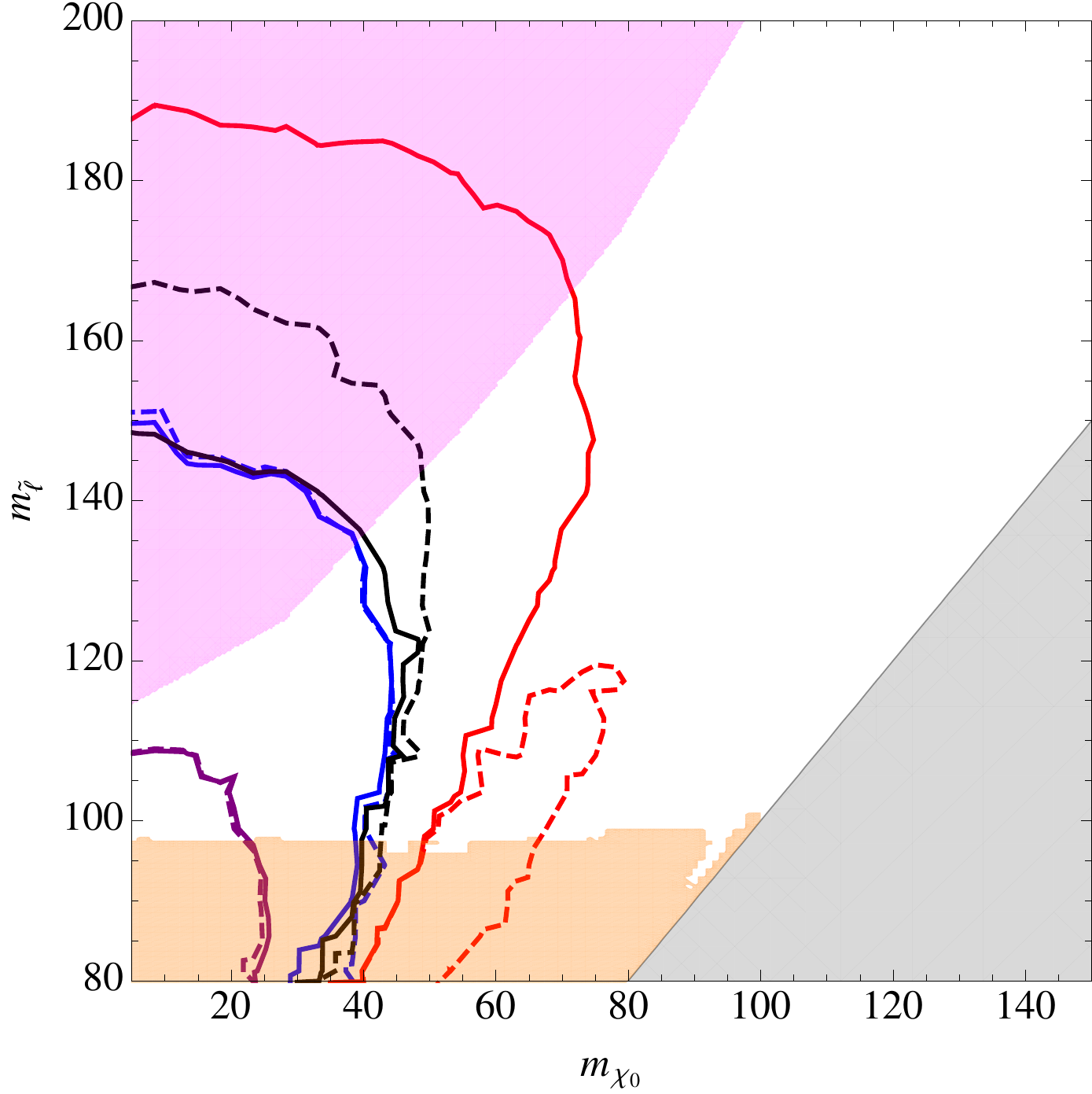} & 
 \includegraphics[width=9cm]{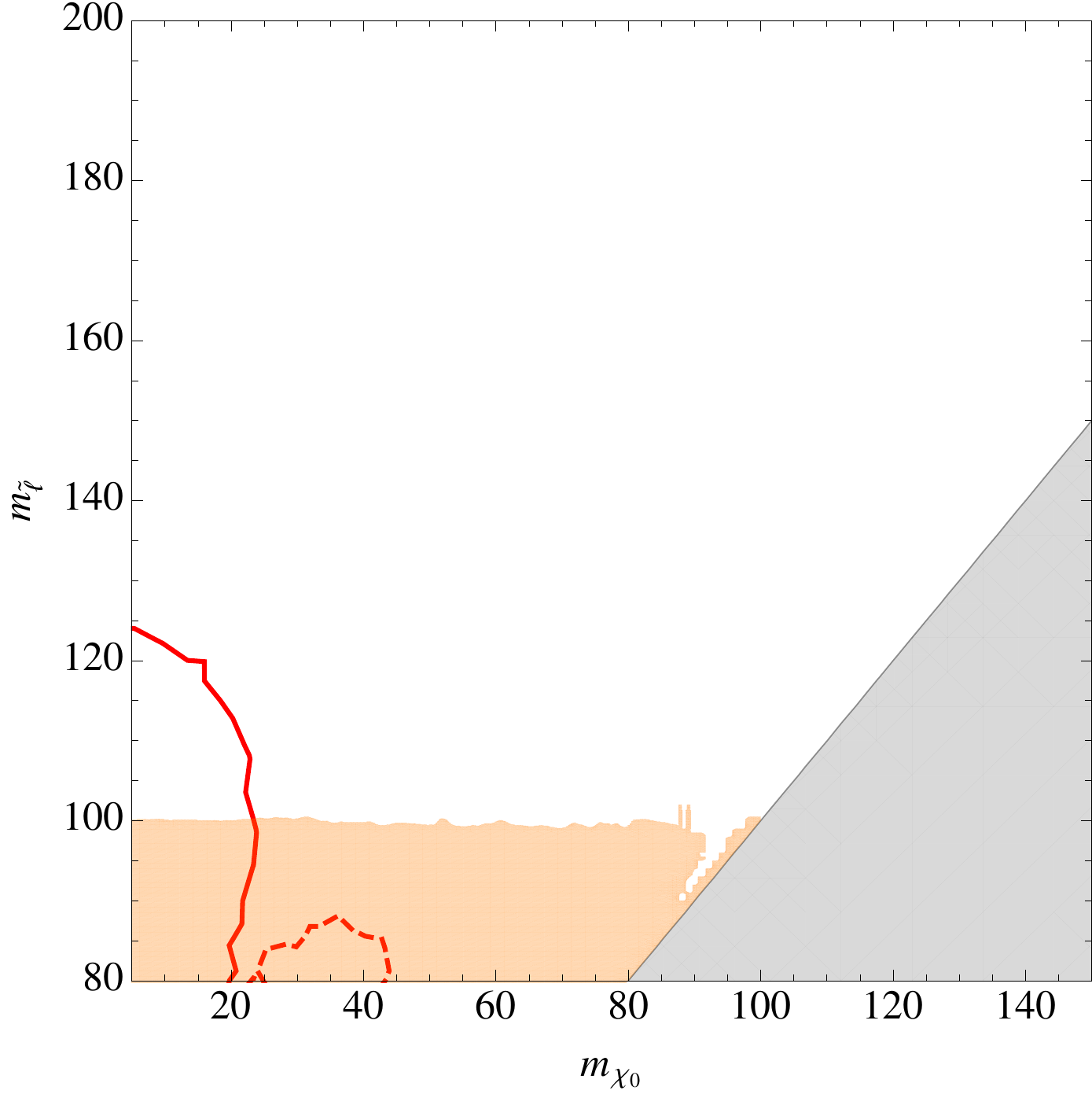} \\
  LH $\tilde e/\tilde \mu \rightarrow \ell + \tilde \chi^0_1$ 
  & 
   RH $\tilde e/\tilde \mu \rightarrow \ell + \tilde \chi^0_1$ \vspace{2mm}
 \end{tabular}
\hspace*{-1cm}
 \begin{tabular}{c}
 \includegraphics[width=9cm]{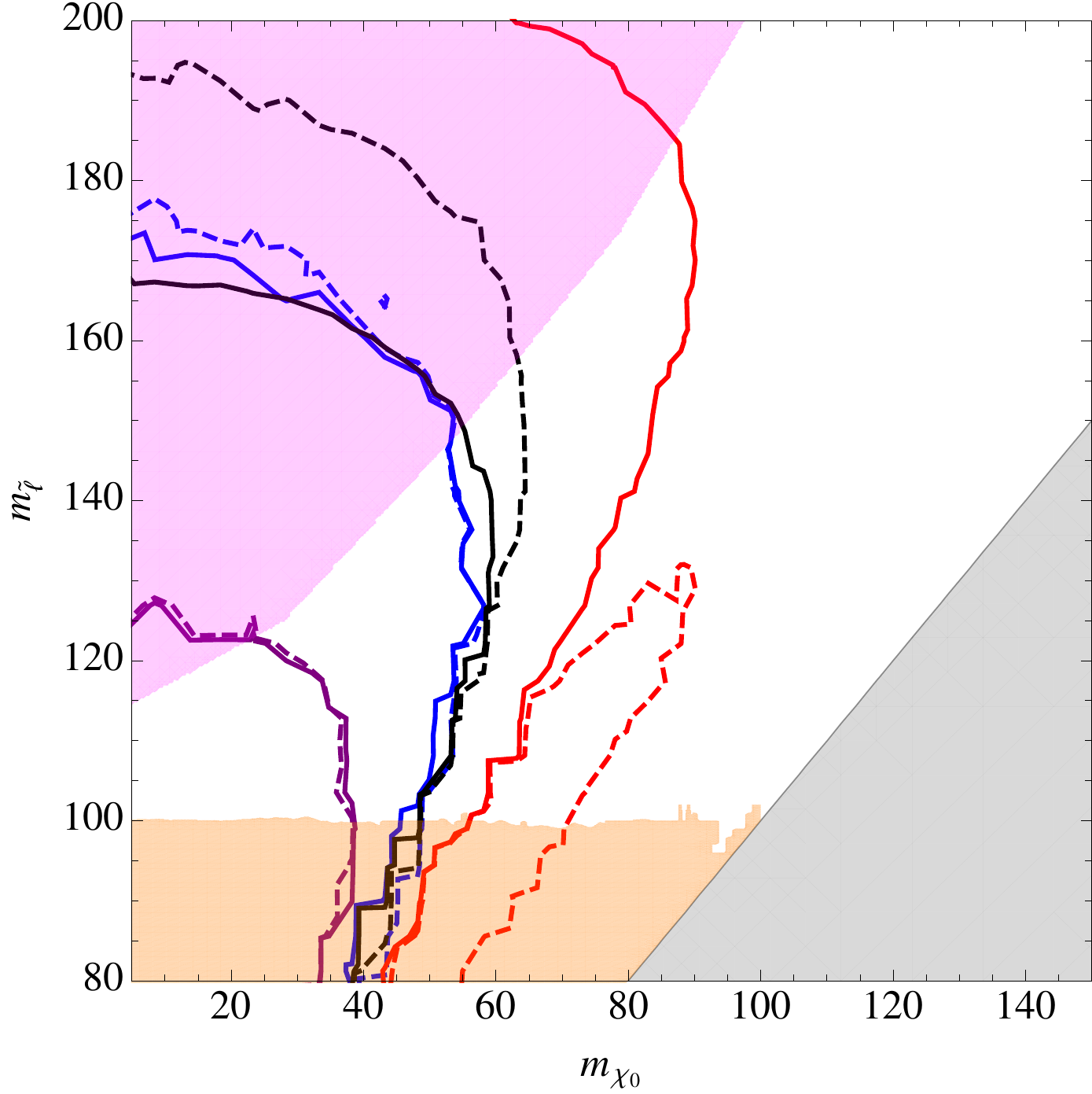} \\ 
  LH + RH $\tilde e/\tilde \mu \rightarrow \ell + \tilde \chi^0_1$
 \end{tabular}\vspace*{-3mm}
    \caption{95\% Exclusions in the neutralino-slepton mass plane for degenerate $\tilde e, \tilde \mu$ decaying to $e/\mu + \tilde \chi^0_1$. Magenta regions are excluded by the CMS $9 \mathrm{fb}^{-1}$ LHC8 slepton search \cite{cmssleptons} (see text footnote). Orange regions are excluded by LEP \cite{LEPlimits}. The regions below the Purple (ATLAS LHC7 \cite{atlas7ww}, Blue (CMS LHC7 \cite{cms7ww}), Red (CMS LHC8 \cite{cms8ww}) and Black (combined) lines are new exclusions we obtained from the respective \ww measurements. Solid (dashed) lines represent limits obtained by (not) renormalizing the SM expectation in all kinematic distributions to match the SM + BSM normalization to data.  The CMS8 \ww measurement was so high that only the region \emph{inside} the red dashed line is not `excluded' when normalization is taken into account.
       }
        \label{f.slepbounds}
   \vspace{-3mm}
\end{center}
\end{figure}

%%%%%%%%%%%%%%%%%%%%%%%%%%%%%%%%%%%%%%%%%%%%%%%%%%%%%%%%%%%%%%%%%%%%%%
%%%%%%%%%%%%%%%%%%%%%%%%%%%%%%%%%%%%%%%%%%%%%%%%%%%%%%%%%%%%%%%%%%%%%%%
\section{Hints of New Physics}
\label{s.newphysics} \setcounter{equation}{0} \setcounter{footnote}{0}
%%%%%%%%%%%%%%%%%%%%%%%%%%%%%%%%%%%%%%%%%%%%%%%%%%%%%%%%%%%%%%%%%%%%%%
%%%%%%%%%%%%%%%%%%%%%%%%%%%%%%%%%%%%%%%%%%%%%%%%%%%%%%%%%%%%%%%%%%%%%

We will now consider a simplified model with  light sleptons and bino dark matter, realized within the MSSM, that can improve agreement with the \ww measurement while also accounting for a range of other anomalies. 

The basic parameter space of our scenario is the $(M_{\mathrm{bino}}, M_{\mathrm{slepton}})$-plane, where $M_{bino} \equiv M_{\tilde \chi^0_1}$ and $M_{slepton} \equiv M_{\tilde e_R} \approx M_{\tilde e_L} \approx M_{\tilde \mu_R} \approx M_{\tilde \mu_L}$ (all within about a GeV of each other) for universal slepton soft masses, which we will mean to also imply $m_{\tilde \ell_L} = m_{\tilde \ell_R}$ unless otherwise stated. Assuming the squarks and gluinos to be above $\sim 500 \gev$ or so along with the heavy higgs scalars (decoupling limit), the remaining relevant parameters are $\mu, \tan \beta, M_2$ and $A_\tau$.

We first show in \ssref{sleptons} that all the \ww measurements prefer a particular region of the $(M_{\mathrm{bino}}, M_{\mathrm{slepton}})$-plane. This is completely independent of $\mu, \tan \beta, M_2$ and $A_\tau$, and depends only on the kinematics of selectron and smuon decay. However, since very light sleptons are preferred, slepton soft mass universality and LEP bounds on the stau mass prefer $A_\tau \sim 0$.

Bino-like dark-matter can have sufficient annihilation cross section through $t$-channel slepton exchange to obtain the correct relic density, while higgs-mediated direct detection is automatically below current bounds but within reach of the next generation of experiments. 
In \ssref{dm} we show that, for wide ranges of the other parameters, the \ww-preferred region and the region of correct relic density intersect in the $(M_{\mathrm{bino}}, M_{\mathrm{slepton}})$-plane. The most important parameters here are $\mu$ and $\tan \beta$, since $A_\tau$ is preferred to be small (see above) and $M_2$ has little effect on the properties of a bino-like LSP. We also explore the departures from slepton soft mass universality to achieve light highly mixed staus and raise $\mathrm{Br}(h\rightarrow \gamma \gamma)$, but we find that a tension between the diphoton rate and the dark matter relic density limits the size of the enhancement to about $\sim 15\%$.

 In \ssref{gm2} we demonstrate how smuon-bino loops can account for the 3 $\sigma$ deviation between observation and SM expectation of the muon anomalous $(g-2)$. This contribution depends, apart from $(M_{\mathrm{bino}}, M_{\mathrm{slepton}})$, mostly on $\mu$ and $\tan \beta$, and we find that our scenario naturally generates a correct size contribution. 
 
We explore the consequences of the slepton soft mass non-universality that is required to moderately enhance $\mathrm{Br}(h\rightarrow \gamma \gamma)$ while achieving the correct dark matter relic density in  \ssref{hgaga}. The resulting lepton-flavor violating operators are tightly constrained, and we show that our scenario would still be well within bounds.

The main success of our scenario is that light sleptons and bino dark matter explain the \ww excess as well as the measured deviation in the muon anomalous $(g-2)$ while producing the correct dark matte relic density. This is shown in \ssref{results}.

%%%%%%%%%%%%%%%%%%%%%%%%%%%%%%%%%%%%%%%%%%%%%%%%%%%%%%%%%%%%%%%%%%%%%%%
\subsection{Sleptons in \ww}
\label{ss.sleptons}
%%%%%%%%%%%%%%%%%%%%%%%%%%%%%%%%%%%%%%%%%%%%%%%%%%%%%%%%%%%%%%%%%%%%%%

\begin{figure}[htbp] %  figure placement: here, top, bottom, or page
\vspace*{-1.1cm}
\begin{center}
\hspace*{-1.5cm}
 \begin{tabular}{cc}
 \includegraphics[width=9cm]{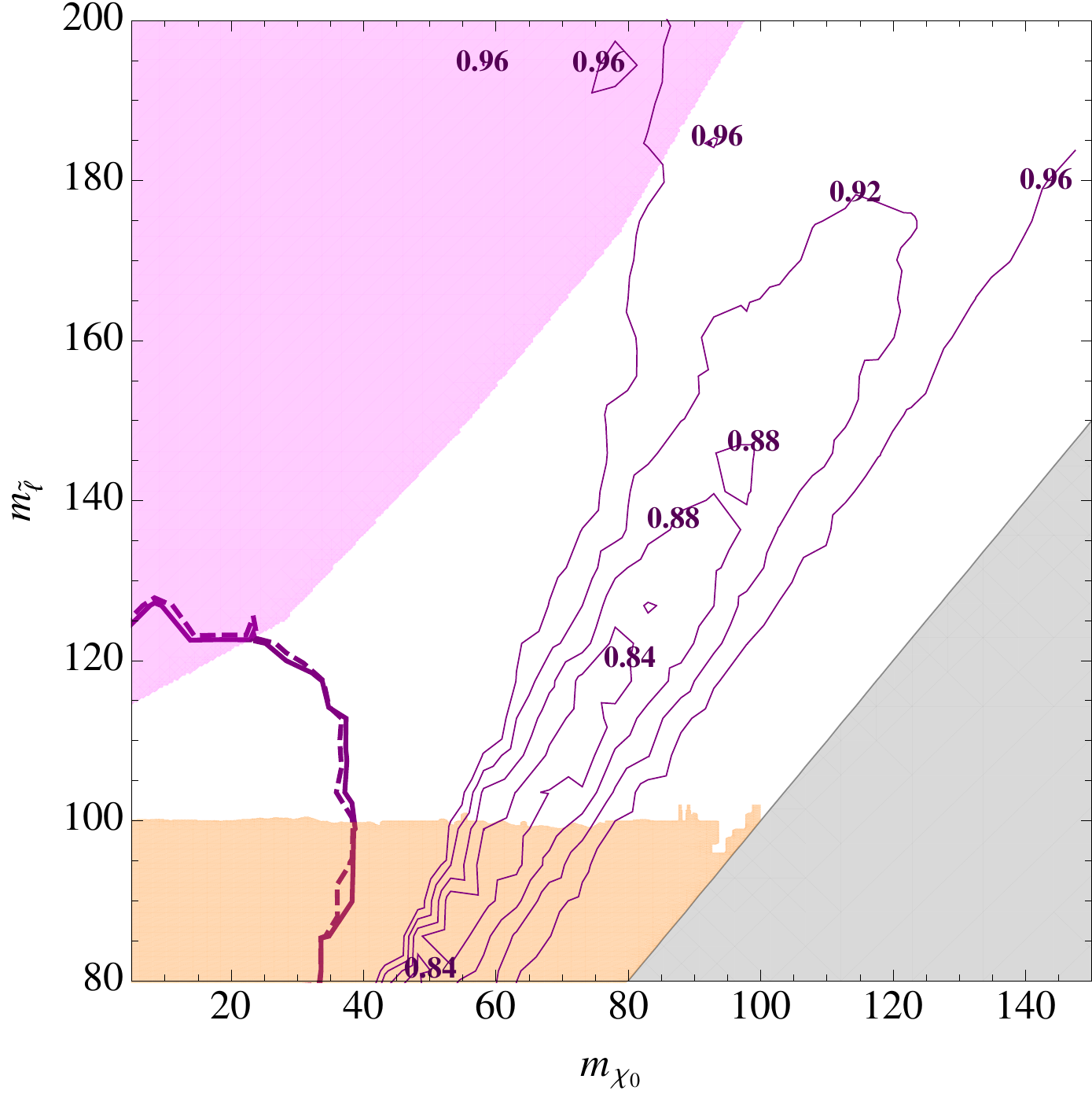} & 
 \includegraphics[width=9cm]{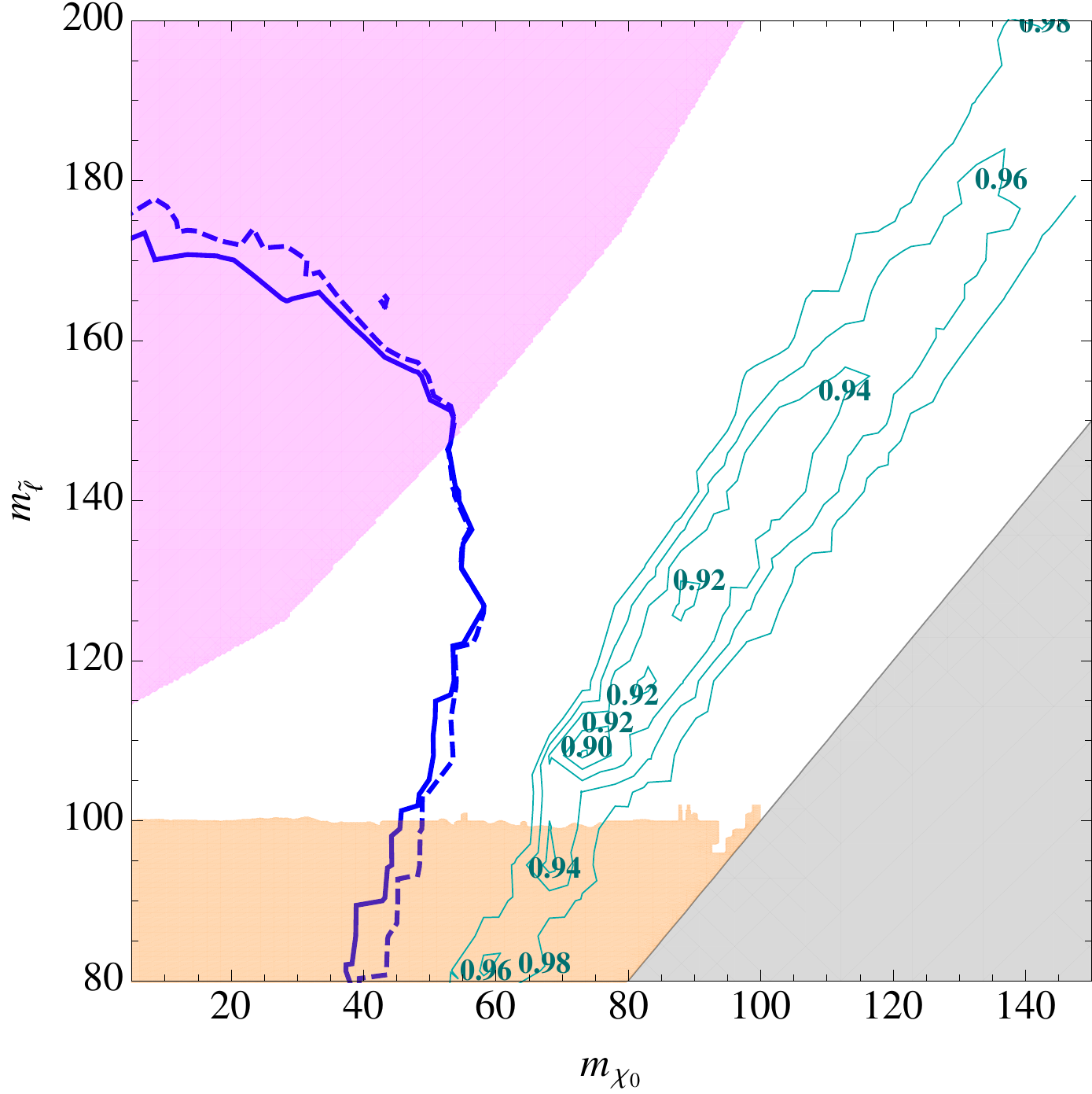} \\ (a) ATLAS LHC7 & (b) CMS LHC7 \vspace{2mm}
 \end{tabular}
 \hspace*{-1.5cm}
 \begin{tabular}{cc}
 \includegraphics[width=9cm]{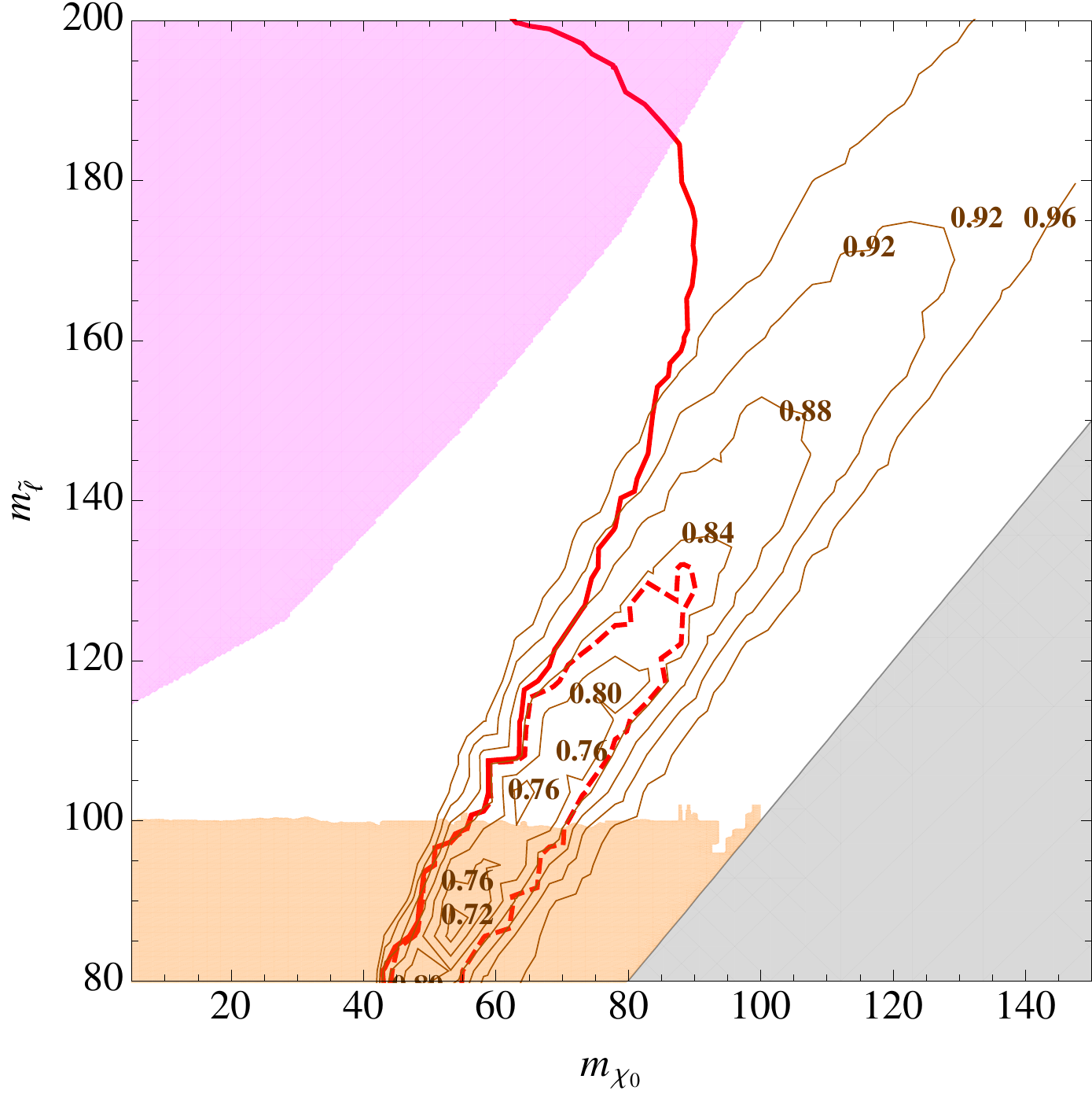} & 
 \includegraphics[width=9cm]{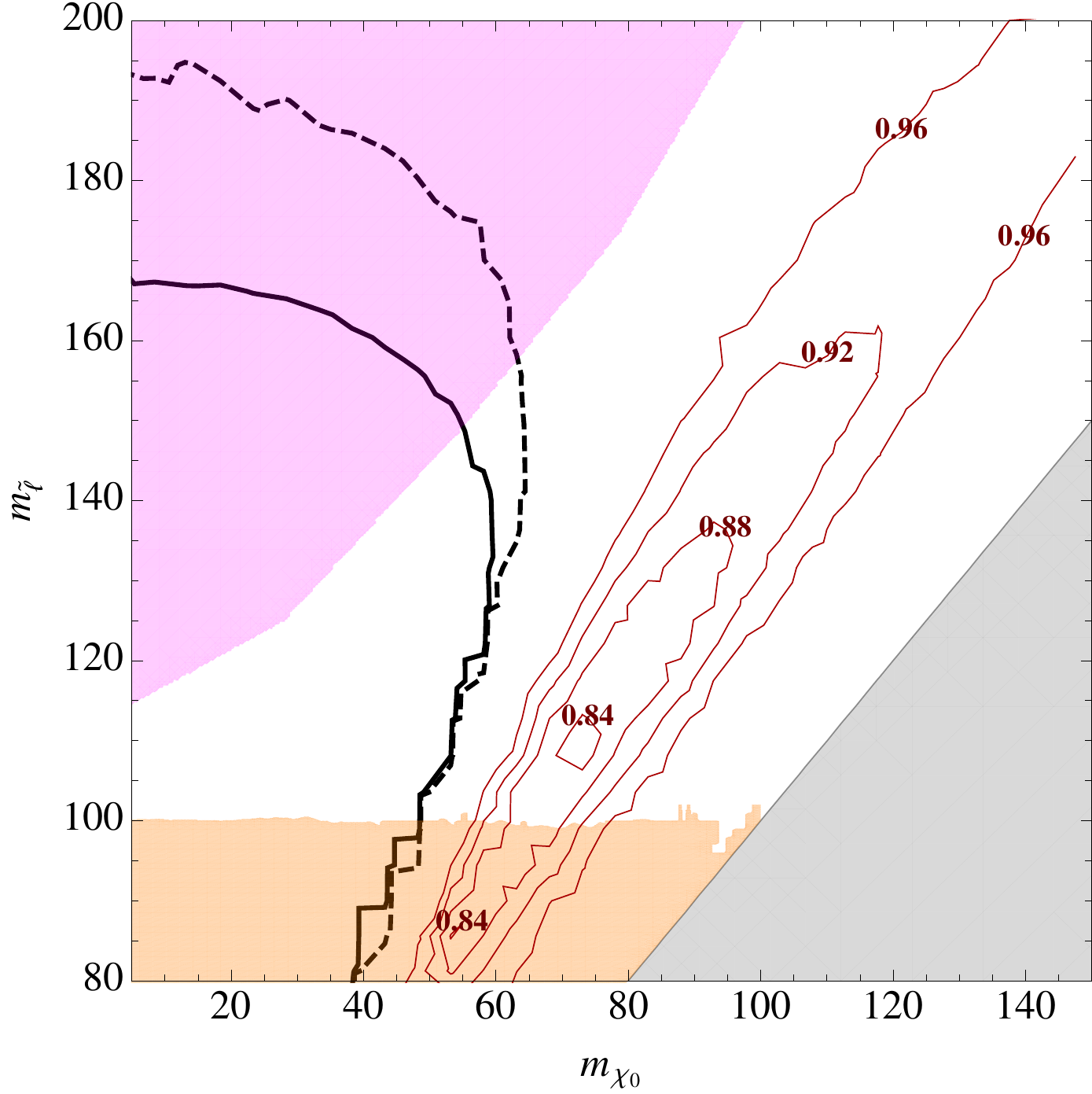} \\ (c) CMS LHC8 & (d) Combined \vspace{2mm}
 \end{tabular} 
    \caption{95\% Exclusions in the neutralino-slepton mass plane for degenerate $\tilde e_L, \tilde e_R, \tilde \mu_L, \tilde \mu_R$ decaying to $e/\mu + \tilde \chi^0_1$. Same color coding as \fref{slepbounds}, but now we also show contours of $r_{\chi^2} = \chi^2_{\mathrm{SM + sleptons}}/\chi^2_{SM}$ where $r_{\chi^2} < 1$, i.e. the slepton contribution improves the fit to data. The overall most preferred point is $m_{\tilde \ell} \approx 120 \gev$, $m_{\tilde \chi^0_1} \approx 80 \gev$.
    }
        \label{f.chi2ratioww}
\end{center}
\end{figure}

Sleptons have a much lower pair production cross section than Charginos~\cite{Curtin:2012nn} or Stops~\cite{Rolbiecki:2013fia}. As a result, even with their higher acceptance in the ATLAS and CMS searches and their $100 \%$ dileptonic branching fraction, they have to be very light in order to meaningfully contribute to the \ww cross section. To make this statement more quantitative it is instructive to revisit the slepton bounds we derived in \sref{bounds}.  

\fref{chi2ratioww} shows the bounds for the RH+LH slepton scenario for each experiment. In addition to the mass bounds, we also show regions where sleptons \emph{improve} the fit to data, as indicated by contours of  $r_{\chi^2} = \chi^2_{\mathrm{SM + sleptons}}/\chi^2_{SM}$ where $r_{\chi^2} < 1$. In those regions the \ww measurement obviously cannot set a bound. Note that \emph{all} the measurements separately or combined prefer the same region in the bino-slepton mass plane, defined roughly by $m_{\tilde \ell} - m_{\tilde \chi^0_1} \approx m_W/2$ and $m_{\tilde \ell}  \lsim 150 \gev$, with the most improvement achieved for $m_{\tilde \ell}  \lsim 120 \gev$. This is squarely in the $WW$-like funnel and invisible to dedicated slepton searches.

Some kinematic distributions from the ATLAS7, CMS7 and CMS8 \ww measurements are shown in \fref{diffww}. The slepton contribution for one of the most preferred points, $m_{\tilde \ell} \approx 110 \gev$ and $m_{\tilde \chi^0_1} \approx 70 \gev$, is added. Note how the BSM contribution affects the bulk of the distributions where the experimental excess lies, and not the hard tail where the SM is in good agreement with data (and anomalous triple gauge couplings would contribute). This is contrary to one possible naive expectation, namely that the two-body decay of sleptons to a lepton + MET would produce a much harder spectrum than $WW$ or charginos. The improvement is particularly stark for the CMS8 measurement: the $p$-value of the SM expected kinematic distributions fitting the observed data is $\sim 10^{-3}$, improved to 0.13, 0.15 or 0.57 by adding sleptons, a 125 GeV SM higgs, or both respectively.

It is important to point out that these conclusions do not depend on all the variables of our scenario. In fact, they \emph{only} depend on the slepton and neutralino masses (and the assumption that $m_{\tilde e_L} \approx m_{\tilde e_R}  \approx m_{\tilde \mu_L} \approx m_{\tilde \mu_R}$, and that all $\tilde e/ \tilde \mu$ decay to $e/\mu + \tilde \chi^0_1$). This is because slepton production at the LHC is through an $s$-channel $\gamma^*/Z^*$ with fixed gauge couplings (unlike at LEP where $t$-channel neutralino contributions are important for selectrons). Staus are also unimportant --- they give almost no contribution to the \ww measurement since two taus have a low dileptonic branching fraction of $\sim 10\%$, and the leptons resulting from that three-body decay are so soft (in our regime of interest) that they do not pass the lepton $p_T$ triggers and cuts of the \ww analysis. However, under the assumption of slepton soft mass universality, the light first and second generation sleptons preferred by the \ww measurement prohibit large stau mixing to avoid LEP bounds on $m_{\tilde \tau_1}$ \cite{LEPlimits}. This implies moderately-sized  $|A_\tau - \mu \tan \beta|  \lsim 3 \tev$, pointing to $\tan \beta \sim 5, \mu \sim 500 \gev$ for $A_\tau \sim 0$.

\newcommand{\tilt}[1]{\begin{sideways}#1\end{sideways}}
\begin{figure}[htbp] %  figure placement: here, top, bottom, or page
\vspace*{-1.1cm}
\begin{center}
\hspace*{-2.0cm} \includegraphics[width=21.5cm]{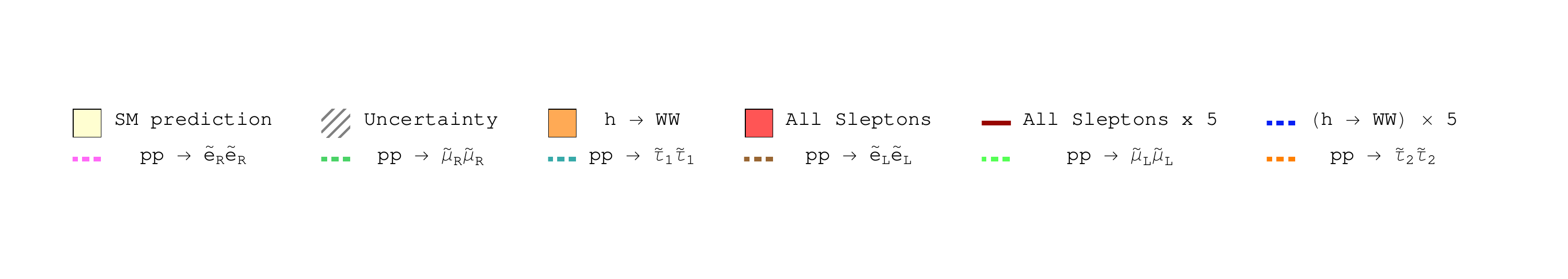}
\vspace*{-16mm}

\hspace*{-1.5cm}
\begin{tabular}{m{9cm}m{9cm}m{1cm}}
\includegraphics[width=9cm]
%ATLAS7
{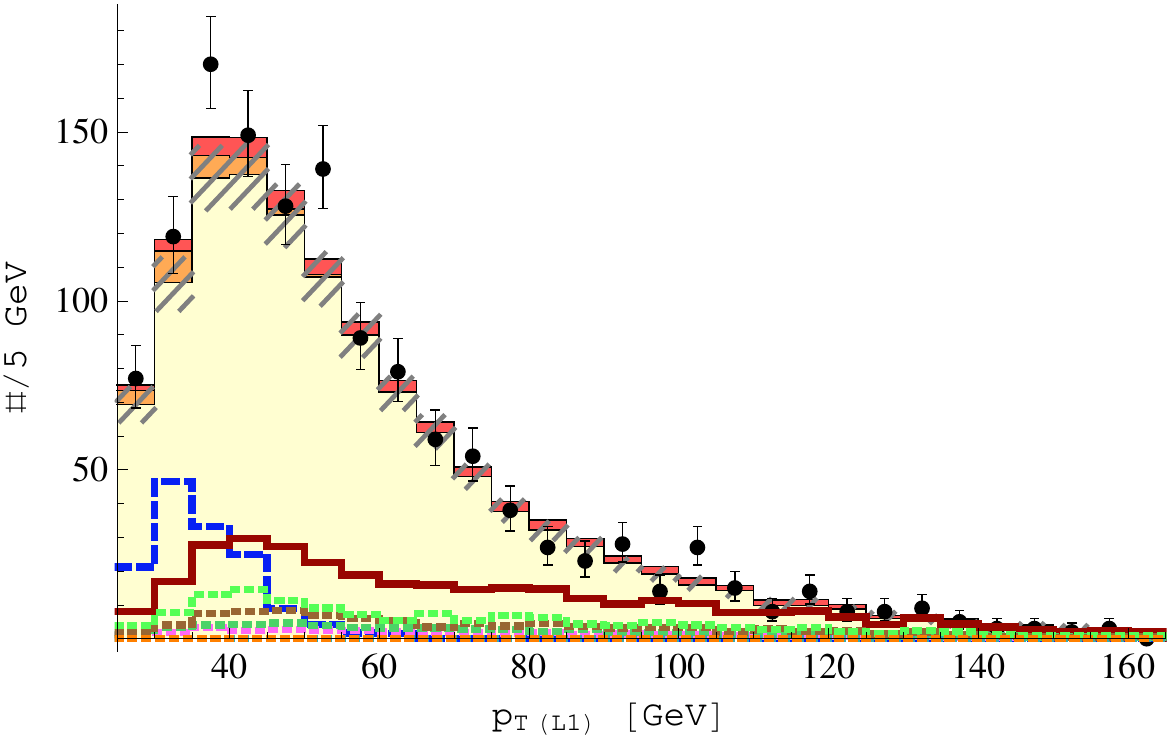}
&
\includegraphics[width=9cm]{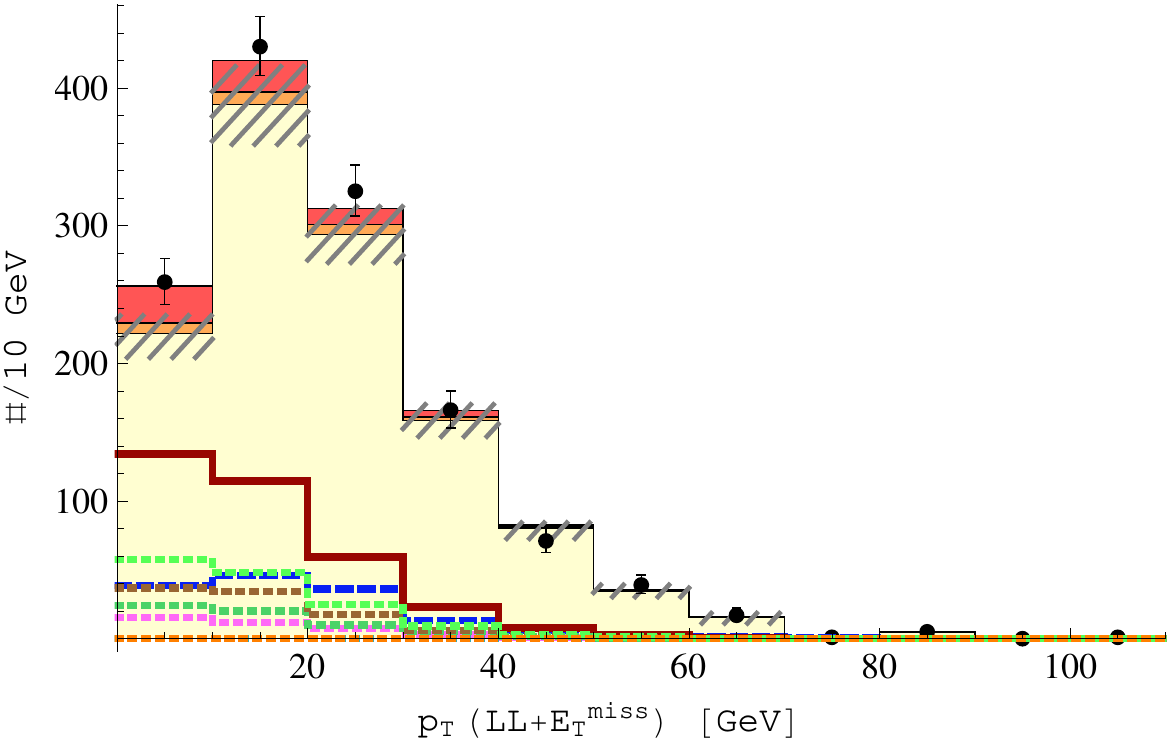}
&
\tilt{\tilt{\tilt{ATLAS LHC7}}}
\\
%CMS7
\includegraphics[width=9cm]{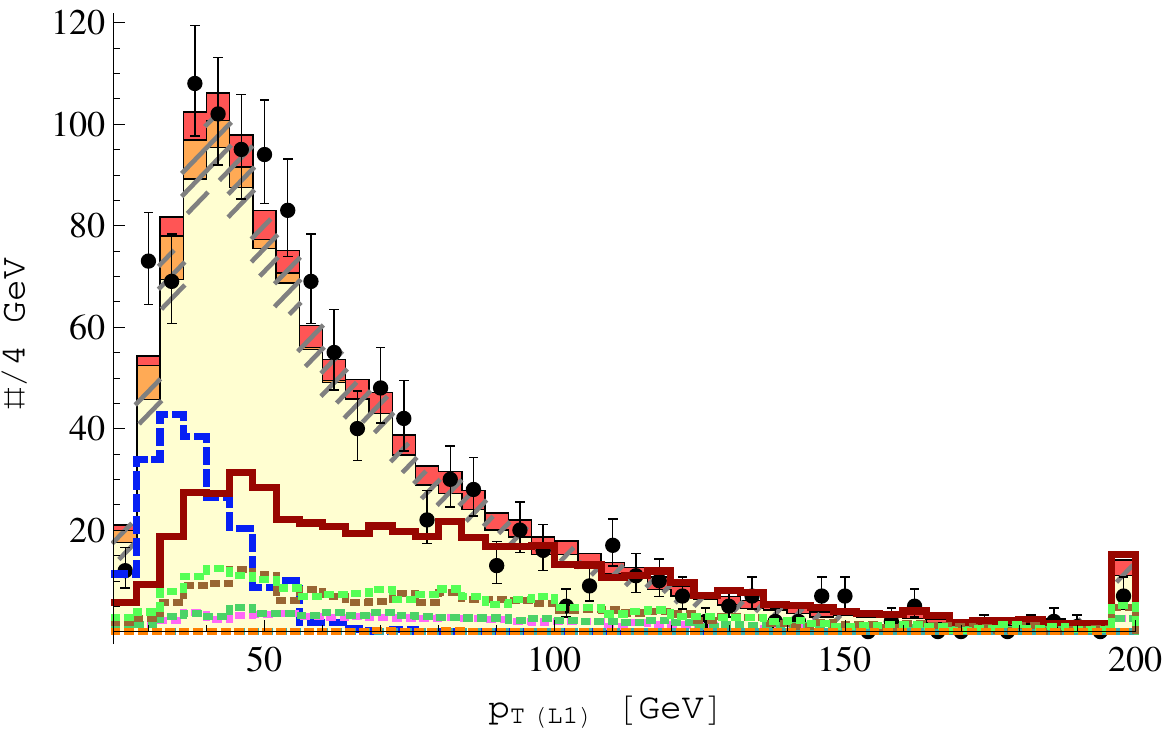}
&
\includegraphics[width=9cm]{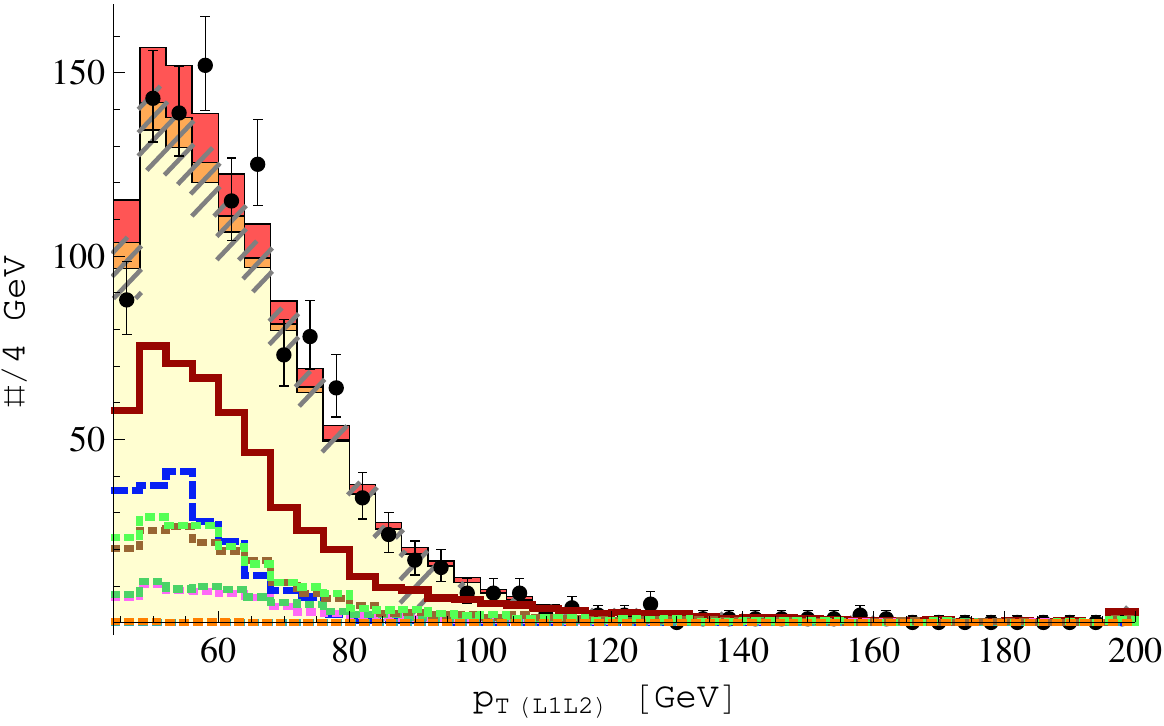}
&
\tilt{\tilt{\tilt{CMS LHC7}}}
\\
%CMS8
\includegraphics[width=9cm]{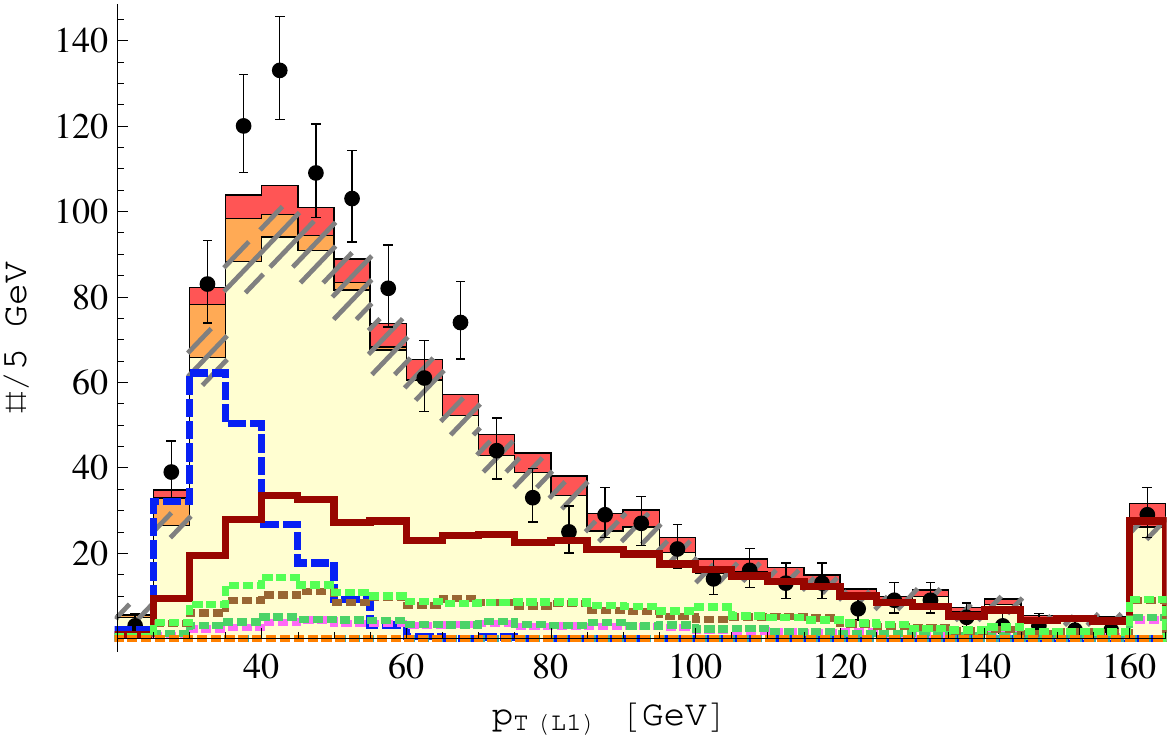}
&
\includegraphics[width=9cm]{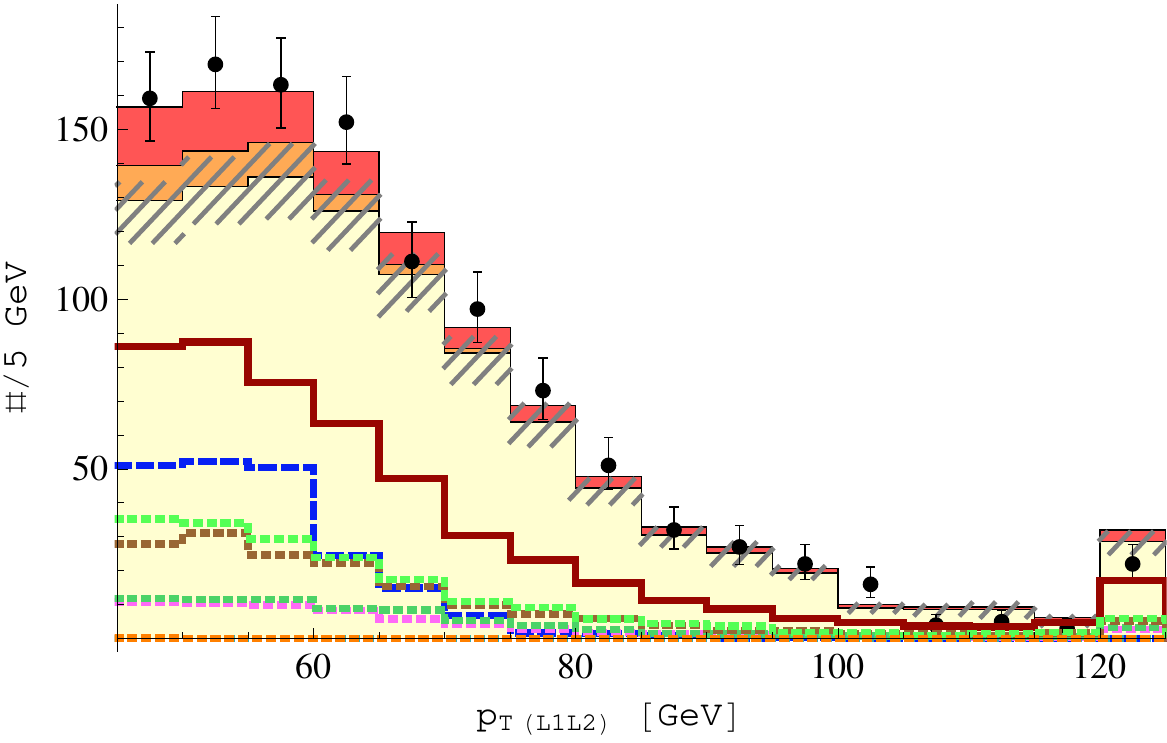}
&
\tilt{\tilt{\tilt{CMS LHC8}}}
\end{tabular}
    \caption{
    Some kinematic Distributions from the \ww measurements done by ATLAS \cite{atlas7ww} and CMS \cite{cms7ww, cms8ww} with slepton contributions for $M_{\mathrm{slepton}} \approx 110 \gev$, $M_{\mathrm{bino}} \approx 70 \gev$ overlaid. The uncertainty refers to the SM prediction. We have also included the effect of a 125 GeV SM higgs, which is a small but non-negligible contribution.
        }
    \label{f.diffww}
\end{center}
\end{figure}

%%%%%%%%%%%%%%%%%%%%%%%%%%%%%%%%%%%%%%%%%%%%%%%%%%%%%%%%%%%%%%%%%%%%%%%
\subsection{Dark Matter}
\label{ss.dm}
%%%%%%%%%%%%%%%%%%%%%%%%%%%%%%%%%%%%%%%%%%%%%%%%%%%%%%%%%%%%%%%%%%%%%%

Dark Matter has long been one of the motivations for physics at the TeV scale, and in particular for Supersymmetry.  While the LHC has ruled out a great deal of SUSY DM parameter space, this is always obliquely through an assumption about charged states, since after all there was a longstanding possibility of heavy Higgsino DM alone which gave both the right relic density and improved unification~\cite{higgsinoDM}~\footnote{This has most recently been ruled out by the HESS measurements~\cite{hess}.}.  The real tension for generic WIMP dark matter are the consistently strong limits placed by direct detection on EW-scale candidates, up to the occasional claims for discovery or odd events.  In particular, if a WIMP carries $SU(2)$ quantum numbers and can interact with nucleons through a $Z$ boson directly it is generally in significant tension with direct detection experiments. This statement is model-independent and does not depend on supersymmetry. 

In the context of the MSSM, one candidate for DM with a small interaction cross section for direct detection experiments is a neutralino that is mostly Bino-like. This is a double-edged sword however, since suppressing the direct-detection cross section can also render the annihilation or co-annihilation cross section very small. This in turn leads to a relic density that is much too high compared to the very precisely measured value given by the Planck satellite \cite{planck}:
\begin{equation}
\Omega_{\mathrm{CDM}} h^2 = 0.1196 \pm 0.0031.
\end{equation}

To generate the correct relic density for a bino-like neutralino, it has been long known that other super partner states are needed to increase the annihilation cross section and achieve the correct relic density  (see e.g. \cite{susyprimer} for a review).   However, it has been pointed out that the window for this is relatively small when resonant annihilation or coannhilation are not relied upon~\cite{welltempered}.  In particular, by analyzing the t-channel annihilation contribution from a single RH slepton (which has the largest coupling to the Bino) the relic density away from those special regions is well described by \cite{welltempered}:
\begin{equation}
\Omega_{\tilde B} h^2 \approx 1.3 \times 10^{-2} \left( \frac{m_{\tilde e_R}}{100 \gev}\right)^2 \frac{(1+r)^4}{r(1+r^2)} \left( 1 + 0.07 \log \frac{\sqrt{r} 100 \gev}{m_{\tilde e_R}}\right),\label{eqn:relic}
\end{equation}
where $ r = {M_1^2}/{m_{\tilde e_R}^2}$.  Without relying upon coannihilation $r\sim 1$ or a particular resonance such as $Z$ or $h$, what naively appears to be a very constrained acceptable region given in~(\ref{eqn:relic}) is precisely the region favored by the \ww cross section measurements which improves the fit to the data.

The dark matter relic density in our scenario depends not just on the  the slepton and bino masses but also on $\mu, M_2, \tan \beta$ and $A_\tau$. This of course just comes from the parametric dependencies of the $\tilde \tau$ mass/coupling, and the mixing of the Bino with the other neutralino gauge eigenstates.    These parameters matter most when there is either a resonant annihilation region or when the $\tilde{\tau}_1$ (or $\tilde \nu_\tau$) becomes lighter and reaches a co-annihilation regime.  Within the simple ansatz of our scenario, the dark matter is always predominantly bino and the  $\tilde{\tau}_1$ mass is never too different from the first and second generation sleptons, the only splitting coming from mixing effects. This automatically leads to the overlap of the preferred \ww collider region with the correct relic abundance in the bino-slepton mass plane.  We have also shown that for moderate ranges of $\mu, M_2, \tan \beta$ and $A_\tau$ our results are also unaffected and the preferred parameter spaces agree.  Finally the spin-independent direct detection cross section in our parameter space is always a factor of a few to $\sim 10$ below the current bounds of $\sim 10^{-45} \mathrm{cm}$ \cite{xenon100} for dark matter masses $M_{\mathrm{bino}} \approx 20 - 200 \gev$. Our scenario therefore avoids all bounds from direct detection, but interestingly enough does predict a potential signal for LUX or XENON1T.  

In \fref{dmabundance} we summarize the relation of the relic density and direct detection limits to our collider parameter space by showing two representative examples of how different $\mu$ and $\tan \beta$ change the DM constraints in our slepton-bino mass plane. The dark matter relic density and direct detection cross section was computed using \texttt{micrOMEGAs 2.4.5} \cite{micromegas}.  It is amusing to note that the resonant annihilation regions that once could be used to accommodate the correct relic density for Bino DM are the same regions that are ruled out by the collider bounds. This is even more obvious in \fref{slepbounds}, where the $m_{\tilde \chi^0_1} \approx m_h/2$ and $m_Z/2$  bands are almost completely excluded by LEP limits and our new slepton exclusion from the CMS8 $WW$ measurement.

\begin{figure}
\begin{center}
\hspace*{-7mm}
\begin{tabular}{cc}
\includegraphics[width=8.5cm]{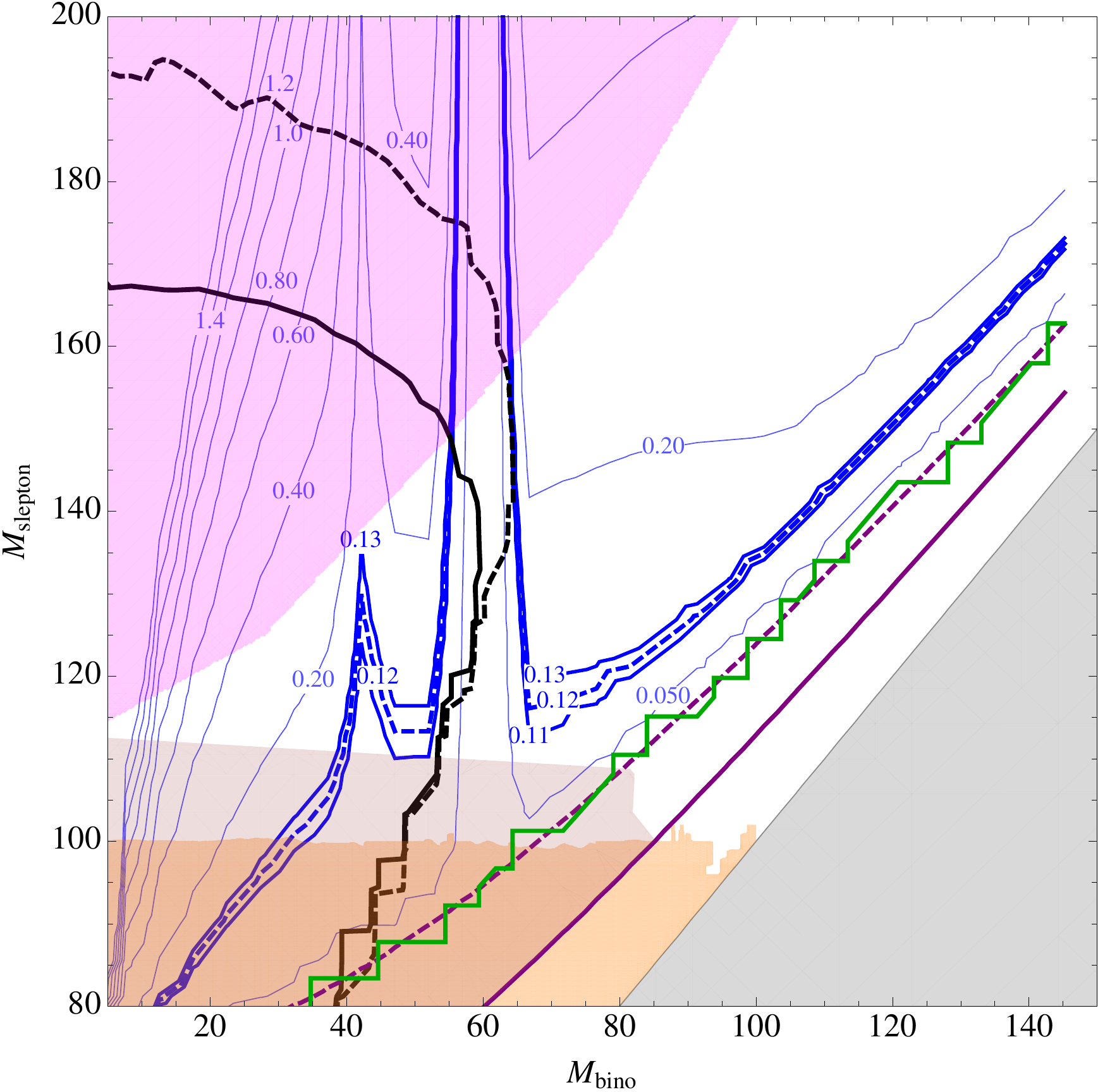}
&
\includegraphics[width=8.5cm]{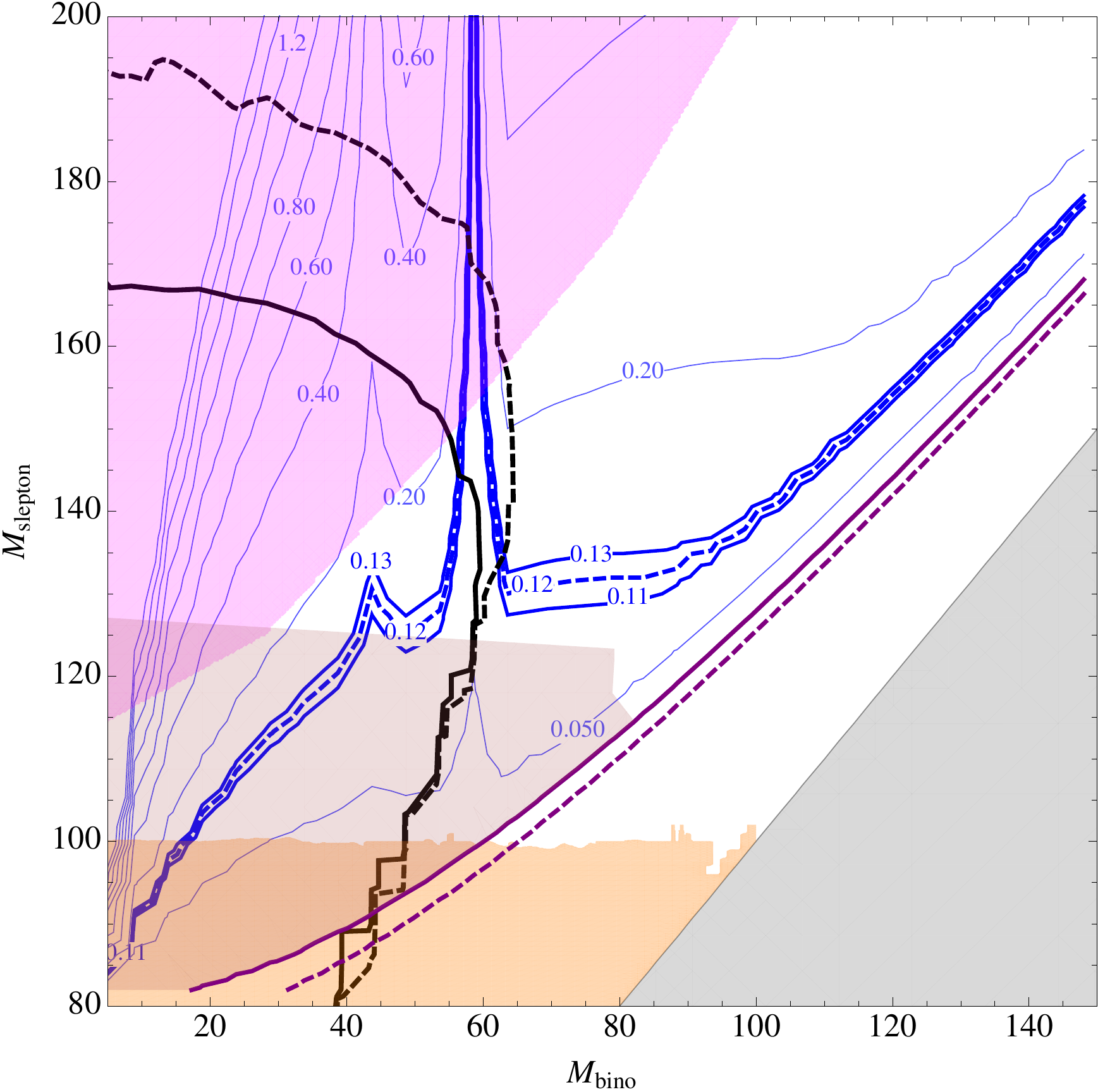}
\\
(a) $\tan \beta = 4, \mu = 400 \gev$ & (b) $\tan \beta = 6, \mu = 600 \gev$ \vspace{4mm} \\
\end{tabular} \vspace*{-8mm}
\end{center}
\caption{The dark matter relic density $\Omega_{\mathrm{CDM}} h^2$ in the $(M_{\mathrm{bino}}, M_{\mathrm{slepton}})$-plane with universal slepton soft masses ($m_{\tilde \ell_L} = m_{\tilde \ell_R}$). The thick dashed and solid lines indicate the best-fit value $\Omega_{\mathrm{CDM}} h^2 = 0.1196$ and the $\pm 3 \times 0.0031 $ values. Grey (Orange) shaded regions are excluded by the LEP bound on $m_{\tilde \tau_1}$ ($m_{\tilde e, \tilde \mu}$)\cite{LEPlimits}. The magenta region is excluded by the CMS slepton search \cite{cmssleptons}, while the black lines indicate our combined slepton bounds from the \ww cross section measurement, see \fref{slepbounds}. Regions below the solid (dashed) purple line have a stau (sneutrino) LSP. Regions below the green line are excluded by the XENON100 direct detection bound \cite{xenon100} on the WIMP-nucleon cross section of $\sim 10^{-45} \mathrm{cm}^2$ for $M_{\mathrm{bino}} \approx 20 - 200 \gev$. $M_2 = 600 \gev$ and $A_\tau = 0$ in this plot.
}
\label{f.dmabundance}
\end{figure}

%DM ABUNDANCE
 
 % slepton soft mass universality:
 %% explain mbino, mslepton dependence
 %% tanbeta, mu dependence

%%%%%%%%%%%%%%%%%% OLD %%%%%%%%%%%%%%%%%%%%%%%%%%%%%

%DD no chance usually if pure bino.
%If we let there be a little mixing (as we'll see g-2 prefers), then the DD will be right on the border of discovery.

%%%%%%%%%%%%%%%%%%%%%%%%%%%%%%%%%%%%%%%%%%%%%%%%%%%%%%%%%%%%%%%%%%%%%%%
\subsection{$(g-2)_\mu$}
\label{ss.gm2}
%%%%%%%%%%%%%%%%%%%%%%%%%%%%%%%%%%%%%%%%%%%%%%%%%%%%%%%%%%%%%%%%%%%%%% 

As it appears in the muon magnetic moment $\vec{M}= g_\mu \frac{e}{2m_\mu} \vec{S}$, the gyromagnetic ratio $g_\mu$ is 2 classically. Since that value receives quantum corrections, the anomalous magnetic moment is defined as $a_\mu \equiv \frac{g_\mu -2}{2}$. This quantity has been the subject of intense theoretical and experimental scrutiny in the last few decades, and the measurement is persistently about $3 \sigma$ higher than the SM prediction \cite{pdg:2012}:
\begin{eqnarray}
\delta a_\mu = a^{exp}_\mu - a^{SM}_\mu = (28.7 \pm 7.98)\times 10^{-10} \label{eqg21}
\end{eqnarray}
The muon anomalous magnetic moment is very sensitive to the existence of BSM charged states that couple to the muon, making it an interesting probe of low-energy supersymmetry. Our scenario features light smuons and binos which, as has been long known, can contribute to $a_\mu$ at one-loop level, see \fref{g2m1}. (The corresponding two-loop contributions are small.) It therefore offers a possible explanation for the observed value of $\delta a_\mu$, with the one-loop contributions explicitly given by \cite{Cho:2011rk}:
\begin{eqnarray}
a_{\mu}\left(\tilde{\chi}^0\right) &=& \frac{-1}{8\pi^2} \sum_{i=1}^{2}\sum_{j=1}^{4}\frac{m_\mu}{m_{\tilde{\mu}_i}} \left\{ \left(|g^{\tilde{\chi}^0_j \mu \tilde{\mu}_i}_L|^2+|g^{\tilde{\chi}^0_j \mu \tilde{\mu}_i}_R|^2\right)\frac{m_\mu}{m_{\tilde{\mu}_i}}G_2\left(\frac{m^2_{\tilde{\chi}^0_j}}{m^2_{\tilde{\mu}_i}}\right)  \right.  \nonumber\\
&+& \left. Re[\left(g^{\tilde{\chi}^0_j \mu \tilde{\mu}_i}_R\right)^* g^{\tilde{\chi}^0_j \mu \tilde{\mu}_i}_L]\frac{m_{\tilde{\chi}^0_j}}{m_{\tilde{\mu}_i}}G_4\left(\frac{m^2_{\tilde{\chi}^0_j}}{m^2_{\tilde{\mu}_i}}\right)
\right\}  \label{e.g2m2} \\
a_{\mu}\left(\tilde{\chi}^-\right) &=& \frac{1}{8\pi^2} \frac{m_\mu}{m_{\tilde{\nu}_\mu}} \sum_{i=1}^{2} \left\{ 
\left(|g^{\tilde{\chi}^-_j \mu \tilde{\nu}_\mu}_L|^2 + |g^{\tilde{\chi}^-_j \mu \tilde{\nu}_\mu}_R|^2 \right)\frac{m_\mu}{m_{\tilde{\nu}_\mu}}G_1\left(\frac{m^2_{\tilde{\chi}^-_j}}{m^2_{\tilde{\nu}_\mu}}\right)  \right. \nonumber\\
&+& \left. Re[\left(g^{\tilde{\chi}^-_j \mu \tilde{\nu}_\mu}_R\right)^* g^{\tilde{\chi}^-_j \mu \tilde{\nu}_\mu}_L] \frac{m_{\tilde{\chi}^-_j}}{m_{\tilde{\nu}_\mu}} G_3\left(\frac{m^2_{\tilde{\chi}^-_j}}{m^2_{\tilde{\nu}_\mu}}\right) \right\} \label{e.g2m1}
\end{eqnarray}

The $G_{i}$ are loop integrals. This formula is convenient for computation but not very illuminating. It is more instructive to examine the contributions in the gauge-eigenstate basis \cite{Cho:2011rk}:\begin{eqnarray}
a_\mu(\tilde{B}, \tilde{\mu}_L- \tilde{\mu}_R) &=& \frac{g^2_Y}{8\pi^2} \frac{m^2_\mu \mu \tan\beta }{M^3_1}F_b\left(\frac{m^2_{\tilde{\mu}_L}}{M^2_1}, \frac{m^2_{\tilde{\mu}_R}}{M^2_1}\right) \label{e.g21}
\\
a_\mu(\tilde{B}-\tilde{H}, \tilde{\mu}_L) &=& \frac{g^2_Y}{16\pi^2} \frac{m^2_\mu M_1 \mu \tan\beta}{m^4_{\tilde{\mu}_L}}F_b\left(\frac{M^2_1}{m^2_{\tilde{\mu}_L}},\frac{\mu^2}{m^2_{\tilde{\mu}_L}}\right) \label{e.g22}
\\
a_\mu(\tilde{B}-\tilde{H}, \tilde{\mu}_R) &=& -\frac{g^2_Y}{8\pi^2} \frac{m^2_\mu M_1 \mu \tan\beta}{m^4_{\tilde{\mu}_R}} F_b\left(\frac{M^2_1}{m^2_{\tilde{\mu}_R}},\frac{\mu^2}{m^2_{\tilde{\mu}_R}}\right) \label{e.g23}
\\
a_\mu(\tilde{W}^0-\tilde{H}, \tilde{\mu}_L) &=& -\frac{g^2}{16\pi^2} \frac{m^2_\mu M_2 \mu \tan\beta}{m^4_{\tilde{\mu}_L}} F_b\left(\frac{M^2_2}{m^2_{\tilde{\mu}_L}},\frac{\mu^2}{m^2_{\tilde{\mu}_L}}\right) \label{e.g24}
\\
a_\mu(\tilde{W}^\pm-\tilde{H}, \tilde{\nu}_\mu) &=& \frac{g^2}{8\pi^2} \frac{m^2_\mu M_2 \mu \tan\beta}{m^4_{\tilde{\nu}}} F_a\left(\frac{M^2_2}{m^2_{\tilde{\nu}}},\frac{\mu^2}{m^2_{\tilde{\nu}}}\right) \label{e.g25}
\end{eqnarray}
A dash, as in $\tilde \mu_L- \tilde \mu_R$, indicates a corresponding mixing insertion. First, notice that all contributions are proportional to $\mu \tan \beta$, which is due to smuon mixing\footnote{$A_\mu$ is assumed to be zero.}. (There is additional dependence on $\mu$ in the loop functions $F_{a,b}$ but this comes from higgsino mixing.) The neutralino contributions ($\ref{e.g21} - \ref{e.g24}$) are all neutralino contributions corresponding to \eref{g2m2}. These contributions dominate for bino dark matter and large $M_2$. While this may be intuitively obvious, the prefactor of $M_2$ in \eref{g25} might imply the contribution to grow with chargino mass. However, the loop functions $F_{a,b}(x,)$ decrease with increasing $x,y$, meaning for our realm of interest for our scenario ($M_2 \gsim 200 \gev$) the chargino contribution is smaller than the bino contribution by a factor of $\sim 4-8$.

\begin{figure}
\begin{center}$\hspace{0cm}
\begin{array}{ll}
\includegraphics[width=1.0\textwidth]{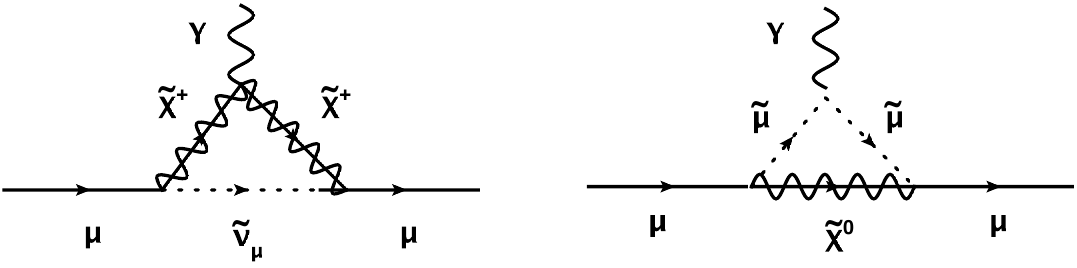}
\end{array}$
\end{center}\caption{$a_\mu$ contributions from neutralinos and charginos in terms of mass eigenstates.}
\label{f.g2m1}
\end{figure}

\fref{gm2plot} shows the $\delta a_\mu$ in the $(M_{\mathrm{bino}},M_\mathrm{slepton})$-plane for different $\mu, \tan \beta$, computed in \texttt{CPsuperH 2.3} \cite{cpsuperh}. Increasing the slepton mass decreases $\delta a_\mu$, but the dependence on $M_{\mathrm{bino}}$ has two different regimes: for small $M_1$ \eref{g21} dominates and $\delta a_\mu$ increases with $M_1$; for large $M_1$, Eqns. (\ref{e.g22}), (\ref{e.g23}) give an overall negative contribution that grows with $M_1$ ($m_{\tilde \mu_L} = m_{\tilde \mu_R}$ in our scenario). This explains the maximum value of $\delta a_\mu$ when $M_{\mathrm{bino}} \sim M_\mathrm{slepton}$. 

Within the regions not yet excluded by slepton bounds, $\delta a_\mu$ and $\Omega_{\mathrm{DM}}$ have very similar scaling with $\mu, \tan \beta$, and there are ranges of both parameters where the best-fit regions for both observables overlap to one sigma. Remarkably, that overlap region also lies in the region preferred by \ww measurements.

 \begin{figure}
\begin{center}
\hspace*{-7mm}
\begin{tabular}{cc}
\includegraphics[width=8.5cm]{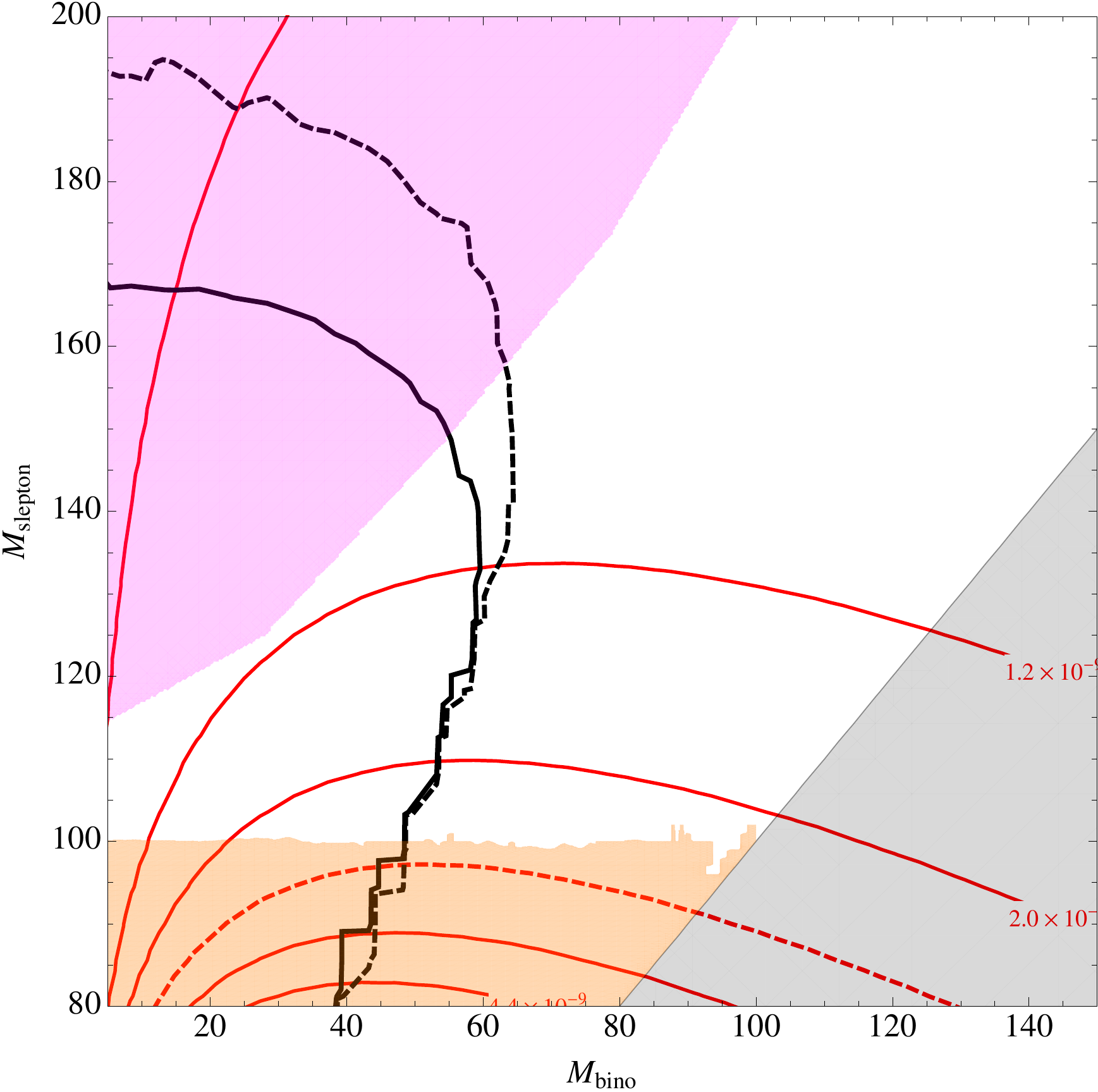}
&
\includegraphics[width=8.5cm]{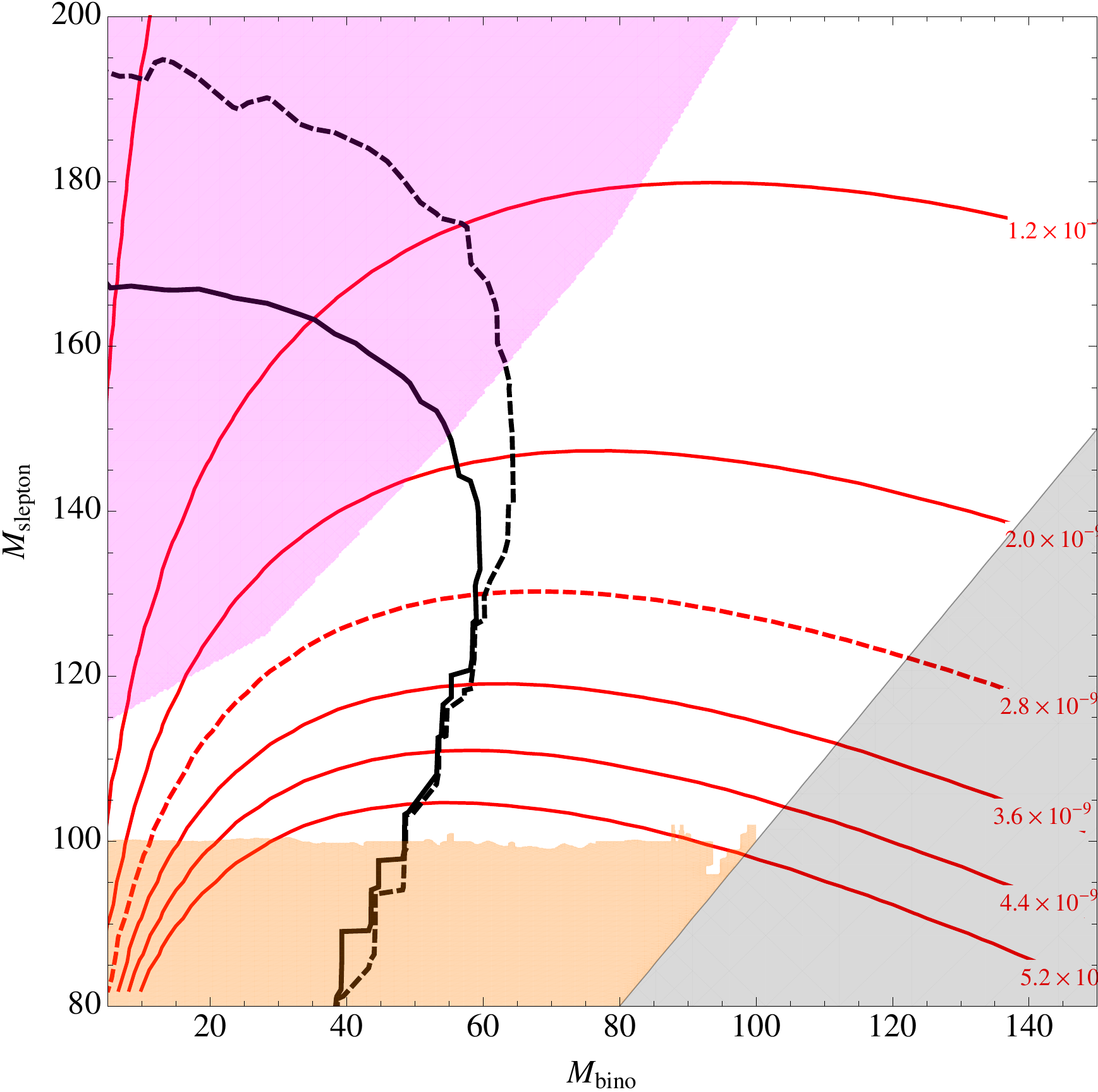}
\\
(a) $\tan \beta = 4, \mu = 400 \gev$ & (b) $\tan \beta = 6, \mu = 600 \gev$
\end{tabular} \vspace*{-8mm}
\end{center}
\caption{Variation of $\delta a_\mu$ in the $(M_\mathrm{bino}, M_\mathrm{slepton})$-plane with universal slepton soft masses ($m_{\tilde \ell_L} = m_{\tilde \ell_R}$). Red lines are contours of $\delta a_\mu$, with the dashed line indicating the experimentally preferred value of $2.87 \times 10^{-9}$ and each contour spacing corresponding to $0.8 \times 10^{-9}$ (one $\sigma$ of the experimental measurement). Grey (Orange) shaded regions are excluded by the LEP bound on $m_{\tilde \tau_1}$ ($m_{\tilde e, \tilde \mu}$)\cite{LEPlimits}. The magenta region is excluded by the CMS slepton search \cite{cmssleptons}, while the black lines indicate our combined slepton bounds from the \ww cross section measurement, see \fref{slepbounds}. 
}\label{f.gm2plot}

\end{figure}

%%%%%%%%%%%%%%%%%%%%%%%%%%%%%%%%%%%%%%%%%%%%%%%%%%%%%%%%%%%%%%%%%%%%%%%
\subsection{$h\rightarrow\gamma\gamma$ and LFU Violation}
\label{ss.hgaga}
%%%%%%%%%%%%%%%%%%%%%%%%%%%%%%%%%%%%%%%%%%%%%%%%%%%%%%%%%%%%%%%%%%%%%%

The LHC has recently discovered a $\approx 125 \gev$ resonance \cite{Higgs125Combined}, properties of which seem consistent with  those of the SM Higgs boson at the $2 \sigma$ level. A mild excess in the diphoton channel, $p p \rightarrow h \rightarrow  \gamma \gamma$, has been reported by the ATLAS experiment \cite{ATLAS:Higgsdiphoton} while a similar excess which had been  reported by the CMS experiment earlier in 2012 \cite{CMS:Higgsdiphoton} has considerably reduced after the new analysis in 2013 \cite{moriond}. Though it is not immediately clear whether the diphoton excess will survive the test of time, it is nevertheless an interesting  possibility that new physics at the electroweak scale can lead to deviations in the $h\rightarrow\gamma\gamma$ effective coupling  through loop induced processes. In particular, the possibility of light staus enhancing the Higgs  diphoton rate has been well-studied  in the literature \cite{LightStaus}. A diphoton rate enhancement of $\sim 50\%$ requires large stau-mixing with $|A_\tau - \mu \tan \beta| \simgt 18 \tev$  and the lightest stau of mass $m_{\tilde{\tau}_1}\sim 90-100 \gev$. This rather narrow \emph{stau window} constrains the soft masses of the third generation sleptons to be $m_{L_3}^2 \approx m_{E_3}^2 \approx m_\tau ( A_\tau - \mu \tan \beta)$. Therefore, a large mass-splitting between the stau eigenstates is induced by electroweak symmetry breaking (EWSB) in this scenario. It also explains why our simple MSSM scenario cannot improve the \ww measurement while enhancing the higgs diphoton rate: under the assumption of slepton soft mass universality, the first and second generation sleptons would be too heavy to significantly influence the measured \ww cross section. 

This motivates us to explore, within our MSSM scenario, a departure from slepton soft mass universality, allowing $m_{\tilde \tau_L} = m_{\tilde \tau_R}$ to differ from the first and second generation  $m_{\tilde \ell} = m_{\tilde \ell_R}$\footnote{The situation is not changed when allowing L and R soft masses to vary independently.}. Since the muon $(g-2)$ is not sensitive to the stau parameters, our task is to understand whether the correct bino relic density is compatible with enhancing $h\rightarrow \gamma \gamma$ by increasing both stau mixing and stau soft masses. 

To answer this question, we investigated the three-dimensional parameter space $(\mu, \tan \beta, X_{\tau})$, where $X_\tau = \frac{1}{m_{\tau}} \{M^2_{\tilde \tau}\}_{LR}$, which is $(A_\tau - \mu \tan \beta)$ at tree-level. (Using $X_\tau$ instead of $A_\tau$ disentangles stau mixing effects from bino-higgsino mixing effects.) We fixed $m_h \approx 125 \gev$, as well as  $M_{\mathrm{bino}} \approx 80 \gev$ and $M_{\mathrm{slepton}} = 110 \gev$ to minimize tension in the $WW$ measurements. To maximally increase the $h \rightarrow \gamma \gamma$ rate we fixed $m_{\tilde \tau_1} = 100 \gev$ by choosing appropriate soft masses $m^2_{\tilde \tau_L} = m^2_{\tilde \tau_R}$ for a given $\mu, \tan \beta$ and $X_\tau$ (equivalently, $A_\tau$). 

Examining the dependence of $\Omega_{\mathrm{DM}}$, $\mathrm{Br}(h \rightarrow \gamma \gamma)$ and $(g-2)_\mu$ across this parameter space, we found that the requirements of correct relic density and significantly enhanced $h \rightarrow \gamma \gamma$ are impossible to satisfy simultaneously. Increasing the diphoton rate requires large stau mixing, which introduces additional diagrammatic contributions to $t$-channel neutralino annihilation and reduces dark matter density. Since the presence of first and second generation sleptons at $\sim 110 \gev$ (from $WW$ measurements) already guarantees a relic density close to $\Omega_{\mathrm{DM}} h^2 \approx 0.1$, introducing mixed staus increases the annihilation cross section beyond acceptable values. This is readily demonstrated in \fref{dmnonuniversalsleptons}(a), which shows the opposite dependency of the relic density and diphoton rate on stau mixing in the above-described scenario for the $\tan \beta = 10$ slice. (With the $X_\tau$ parameterization of stau mixing, the remaining explicit $\tan \beta$ dependence of $\Omega_\mathrm{DM}$ and $\mathrm{Br}(h\rightarrow \gamma \gamma)$ is small.)  In general, requiring correct dark matter relic density limits the maximum diphoton rate enhancement to about 15\% for first and second generation sleptons lighter than 145 GeV. It also requires $\mu$ to be in the few hundred GeV range, as demonstrated by \fref{dmnonuniversalsleptons}(b): increasing $\mu$ decreases the higgsino fraction of the neutralino, which as explained above significantly reduces relic density due to more efficient annihilation, even when stau mixing is kept constant. 

This paints a very clear picture. If the $WW$ measurements are interpreted as implying light first and second generation sleptons near 110 GeV and a neutralino near 80 GeV, the resulting dark matter annihilation is so efficient that both $\mu$ and stau-mixing must be relatively small. The maximum higgs diphoton enhancement that can be achieved is about 15\%, and requires stau soft masses of $\sim 300 \gev$ compared to first and second generation slepton soft masses of $\sim 100 \gev$, as well as first and second generation sleptons slightly heavier ($\sim 120 - 130 \gev$) than what is ideally preferred by \ww measurements.

\begin{figure}
\begin{center}
\hspace*{-4mm}
\begin{tabular}{ccc}
$\mu = 400 \gev$  & & $X_\tau = -1500 \gev$\\
\includegraphics[width=8cm]{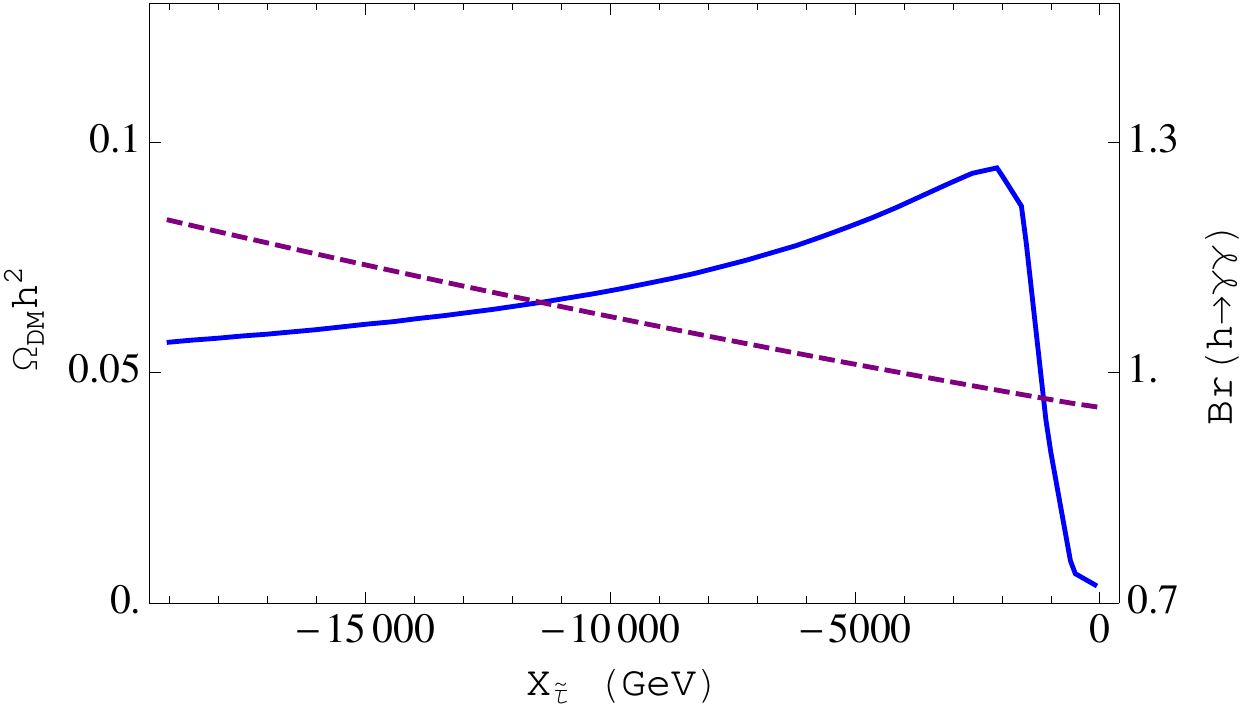}
& \hspace{0mm} &
\includegraphics[width=8cm]{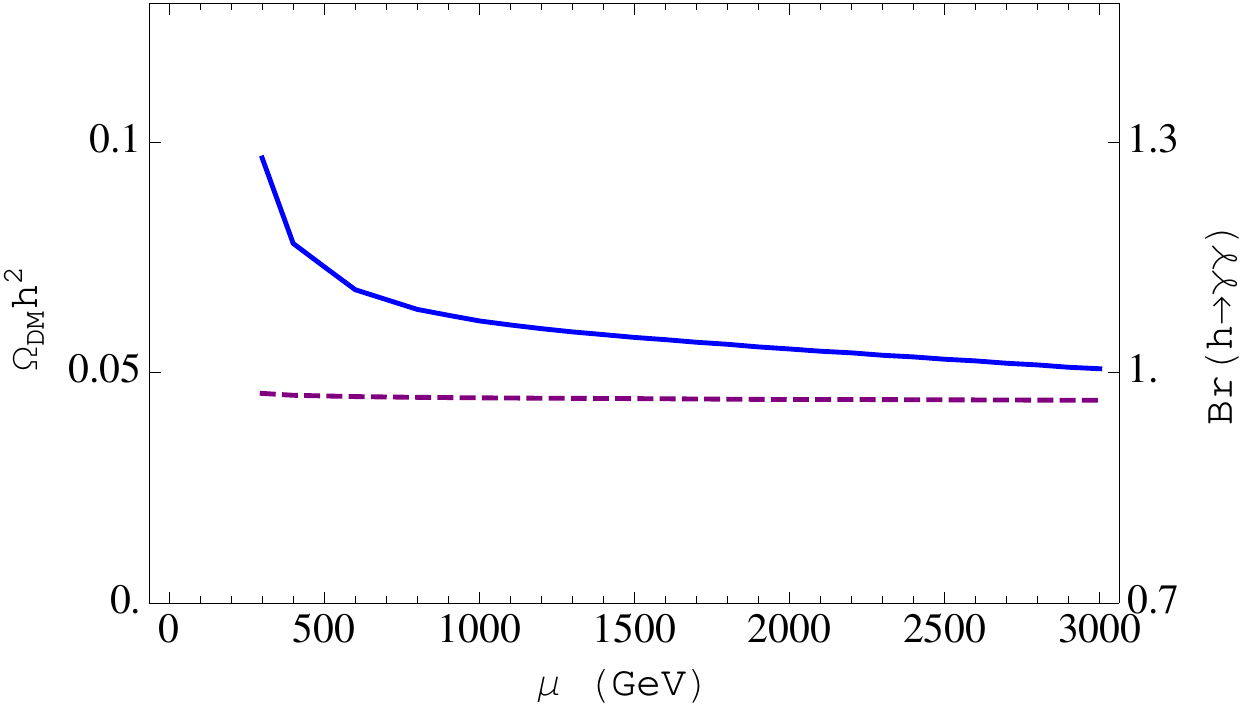}
\\
(a) &&  (b) 
\end{tabular} \vspace*{-8mm}
\end{center}
\caption{
The dark matter relic density (blue solid line) and $\mathrm{Br}(h \rightarrow \gamma \gamma)$ (purple dashed line, normalized to SM) as functions of $X_\tau$ and $\mu$ in the $\tan \beta = 10$ slice. $X_\tau \equiv \frac{1}{m_{\tau}} \{M^2_{\tilde \tau}\}_{LR} = (A_\tau - \mu \tan \beta)$ at tree-level. $M_\mathrm{bino}, M_\mathrm{slepton}$ and $m_{\tilde \tau_1}$ were fixed at 80, 110 and 100 GeV. In (a) relic density increases with decreasing stau mixing. The sudden fall-off near $X_\tau = 0$ is from tau sneutrinos becoming light due to the small stau soft masses with minimal mixing, giving rise to stau-neutralino co-annihilation or even a stau LSP.
}
\label{f.dmnonuniversalsleptons}
\end{figure}

While our scenario cannot explain a 50\%  enhancement, as the measurements of the higgs diphoton rate become more precise a smaller but nonzero enhancement may still be desirable. Since our scenario requires a departure from slepton soft mass universality to achieve a 15\% enhancement, it is prudent to check that the bounds on LFU violation do not exclude this possibility. There are two sources violating lepton flavor universality (LFU) in this scenario : 
 (i) the EWSB induced term proportional to tau Yukawa,  $m_\tau ( A_\tau - \mu \tan \beta)$, which is large in the region where 
 the diphoton rate is enhanced and, 
 (ii) non-degenrate soft SUSY parameters in the slepton sector, which is a necessary condition if both  \ww and Higgs diphoton 
 anomalies are to be reconciled. Even if one is agnostic about the diphoton excess, there are regions in the light slepton parameter
  space where LFU violation can be non-negiligible.\footnote{There is also a possibility of lepton flavour violation from flavour 
  off-diagonal terms in the slepton mass matrix but we do not consider them here since they are known to be small \cite{LFVoffdiag}.}

 \begin{figure}
\begin{center}
\includegraphics[width=9cm]{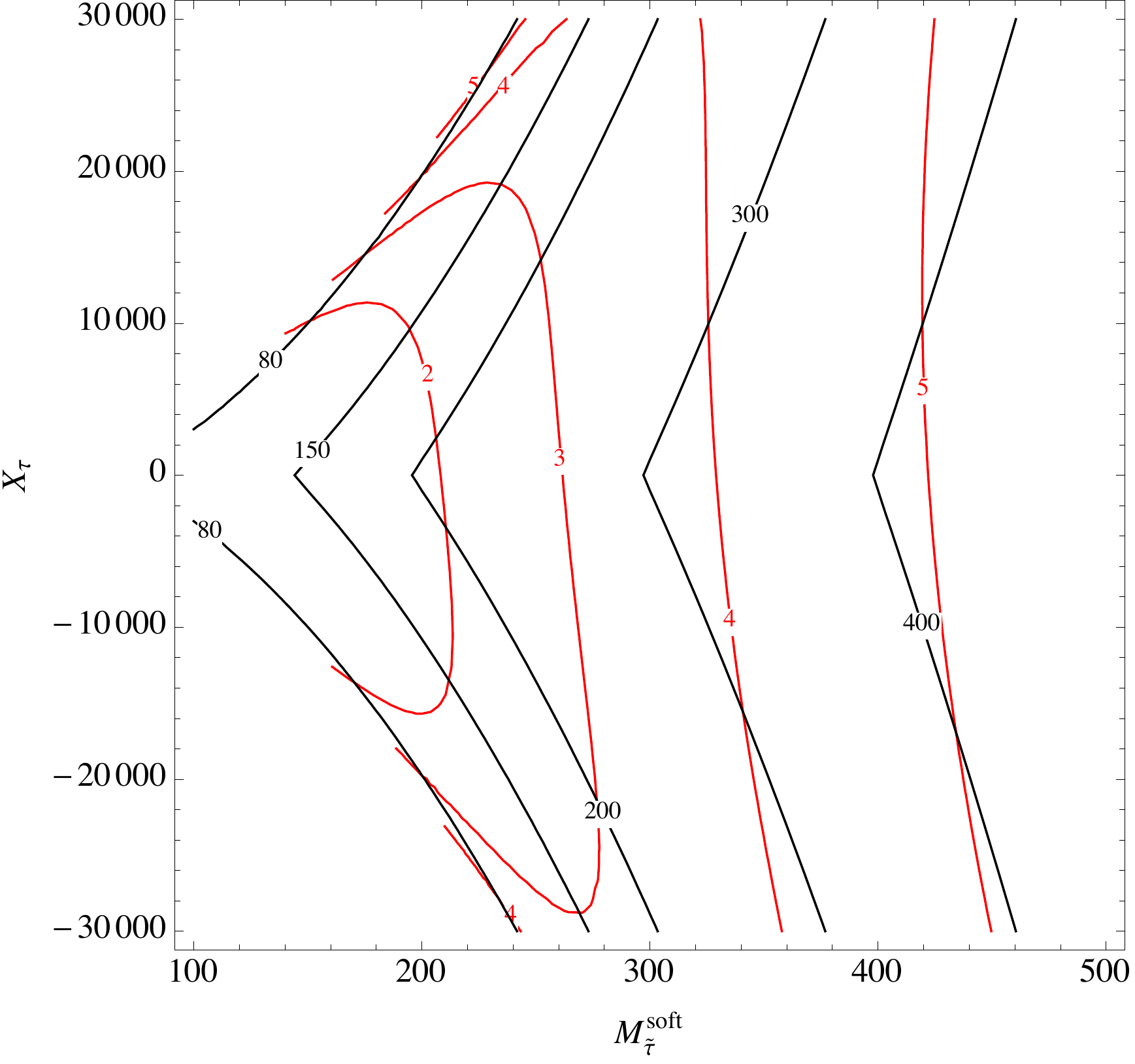}
\end{center}
\caption{
$\Delta r^{\mu/\tau} \times 10^5$ (red contours) as a function of  $X_\tau = \frac{1}{m_{\tau}} \{M^2_{\tilde \tau}\}_{LR}$ (which is $(A_\tau - \mu \tan \beta)$ at tree-level) and $m_{\tilde \tau}^\mathrm{soft}$,  the common $\tilde \tau_L$, $\tilde \tau_R$ soft mass. The gray contours are the $\tilde \tau_1$ mass eigenvalue (all in GeV), and only regions where $m_{\tilde \tau_1} > 80 \gev$ are of interest \cite{LEPlimits}. Across the entire range, $\Delta r^{\mu/\tau}$ is much smaller than the upper experimental bounds, which are $\mathcal{O}(10^{-3}) - \mathcal{O}(10^{-2})$ depending on the process \cite{LFNU:Giudice}. For this plot, the first and second generation slepton soft mass is 100 GeV, $M_1 = 80 \gev$, $\mu = 400 \gev$, $\tan \beta = 4$, $M_2 = 600 \gev$ and $A_\mu = 0$, but changing these parameters will not increase $\Delta r^{\mu/\tau}$  beyond experimental limits.
}
\label{f.lfvplot}
\end{figure}

Bounds on LFU violation  can be parameterized by different effective Fermi constants for different leptons. Considering for example tau vs muon decay, we can define the quantity $\Delta r^{\mu/\tau}$ \cite{LFNU:Giudice} as follows: 
\begin{equation}
\begin{split}
\Delta r^{\mu/\tau} &= \frac{(R_{\mu/\tau})}{(R_{\mu/\tau})_{\text{SM}}}-1 \\
&= \frac{\Gamma(\mu \rightarrow e \nu_\mu \overline{\nu}_e)/\Gamma(\tau \rightarrow e \nu_\tau \overline{\nu}_e)}
{\Gamma(\mu \rightarrow e \nu_\mu \overline{\nu}_e)_{\text{SM}}
/\Gamma(\tau \rightarrow e \nu_\tau \overline{\nu}_e)_{\text{SM}}} - 1 \\
&= \frac{G_\mu^2/G_\tau^2}{(G_\mu^2)_{\text{SM}}/(G_\tau^2)_{\text{SM}}} - 1
\end{split}
\label{e.delta_r}
\end{equation}
where $G_\tau (G_\mu)$ is the Fermi decay constant for tau (muon) decay. Depending on the process for which $\Delta r^{\mu/\tau}$ is evaluated, its absolute value is bounded to be smaller than $\mathcal{O}(10^{-3}) - \mathcal{O}(10^{-2})$.  \cite{LFNU:Giudice, Deltar:Exp}.  

To a good approximation, the theoretical prediction for $\Delta r^{\mu/\tau}$ is process-independent. 
The relation between the measured Fermi 
constant and the $W$ boson mass receives loop corrections depending on the process under consideration and is 
parametrized by $\Delta r^f$ (not to be confused with $\Delta r^{\mu/\tau}$): 
\begin{equation}
G_f = \frac{\pi \alpha}{\sqrt{2} M_W^2 s_w^2} (1 + \Delta r^f) \, , \quad f=\mu,\tau
\end{equation}
where $s_w$ is the sine of the weak mixing angle and $\alpha$ is the electromagnetic constant. Plugging this relation 
back in \eref{delta_r}, we get 
\begin{equation}
\begin{split}
\Delta r^{\mu/\tau} &= \left| \frac{1 + \Delta r^\mu}{1 + \Delta r^\mu_{\text{SM}}} \right|^2  \left| \frac{1 + 
\Delta r^\tau_{\text{SM}}}{1 + \Delta r^\tau} \right|^2 - 1 \\
&\approx 2 (\Delta r^\mu_{\text{SUSY}} - \Delta r^\tau_{\text{SUSY}})
\end{split}
\end{equation}
From the above expression, it is clear that any lepton-universal contributions to $\Delta r^{\mu/\tau}$ cancel out. 
Analytic expressions for supersymmetric contributions to $\Delta r$ are presented in \cite{mssm:deltar} but are too lengthy to reproduce here. The most important input is the splitting between the stau soft masses and the first/second generation soft masses, and the stau mixing $X_\tau$, and includes both sources of LFU violation mentioned above. \fref{lfvplot} shows $\Delta r^{\mu/\tau}$ for representative choices of parameters, and it is always orders of magnitude below experimental bounds for the relevant parameter ranges. Therefore, while our scenario does not naturally account for a $h \rightarrow \gamma \gamma$ enhancement, a moderate enhancement of $\sim 15 \%$ may be accommodated.

%%%%%%%%%%%%%%%%%%%%%%%%%%%%%%%%%%%%%%%%%%%%%%%%%%%%%%%%%%%%%%%%%%%%%%%
\subsection{Summary}
\label{ss.results}
%%%%%%%%%%%%%%%%%%%%%%%%%%%%%%%%%%%%%%%%%%%%%%%%%%%%%%%%%%%%%%%%%%%%%%

 \begin{figure}
 \vspace*{-20mm}
\begin{center}
\hspace*{-7mm}
\begin{tabular}{cc}
\includegraphics[width=8.5cm]{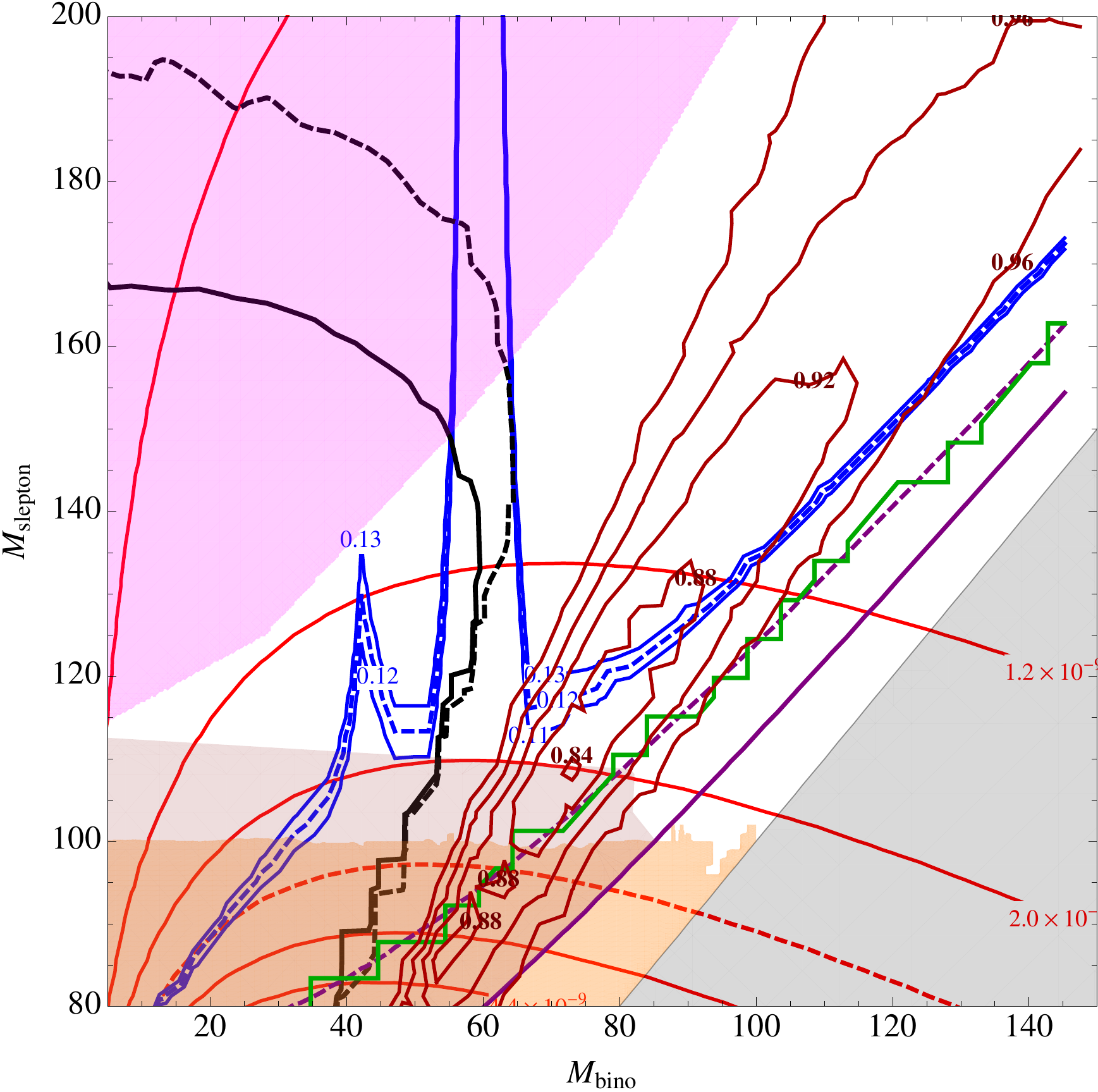}
&
\includegraphics[width=8.5cm]{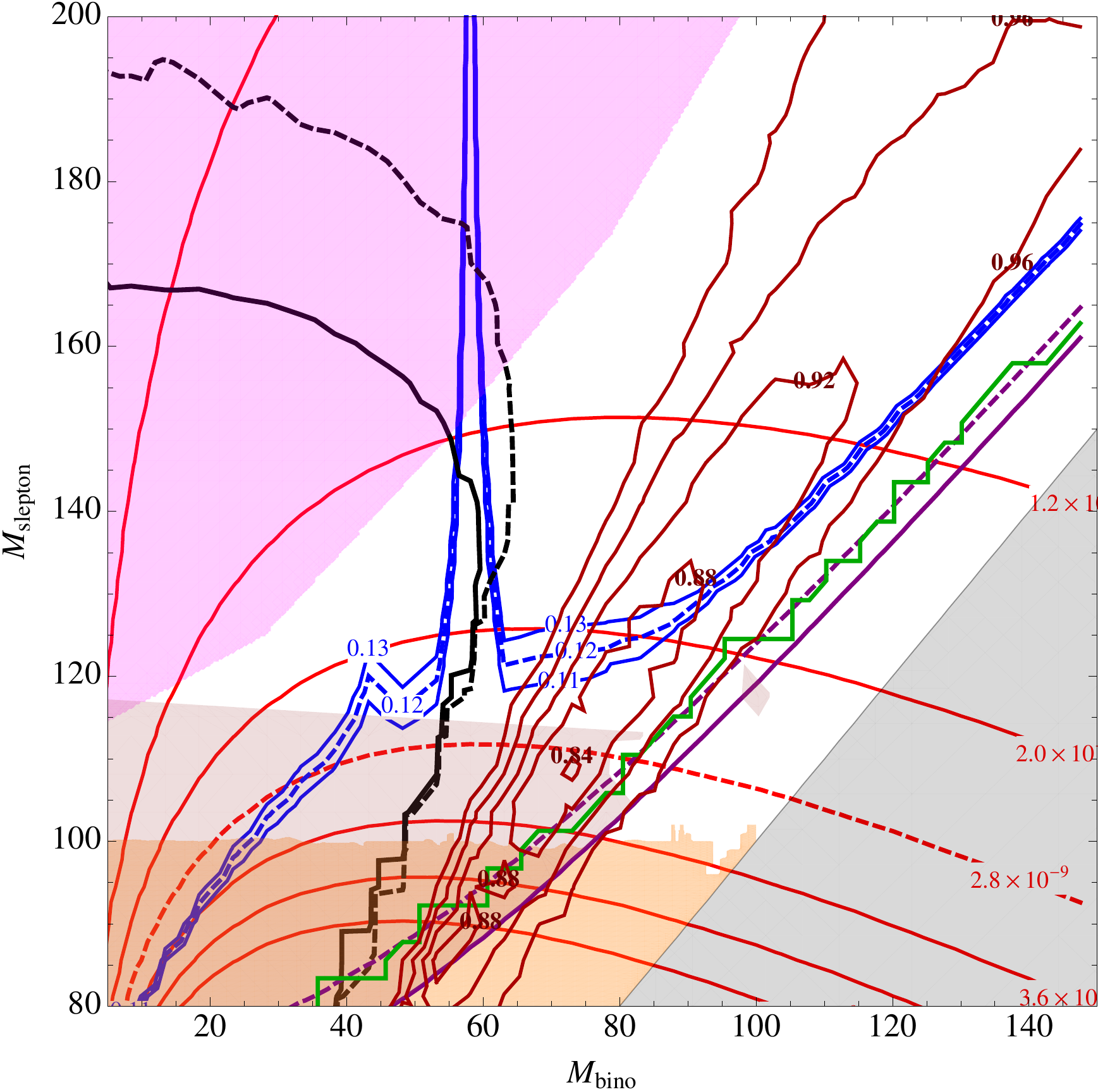}
\\
(a) $\tan \beta = 4, \mu = 400 \gev$ & (b) $\tan \beta = 4, \mu = 600 \gev$ \\
\includegraphics[width=8.5cm]{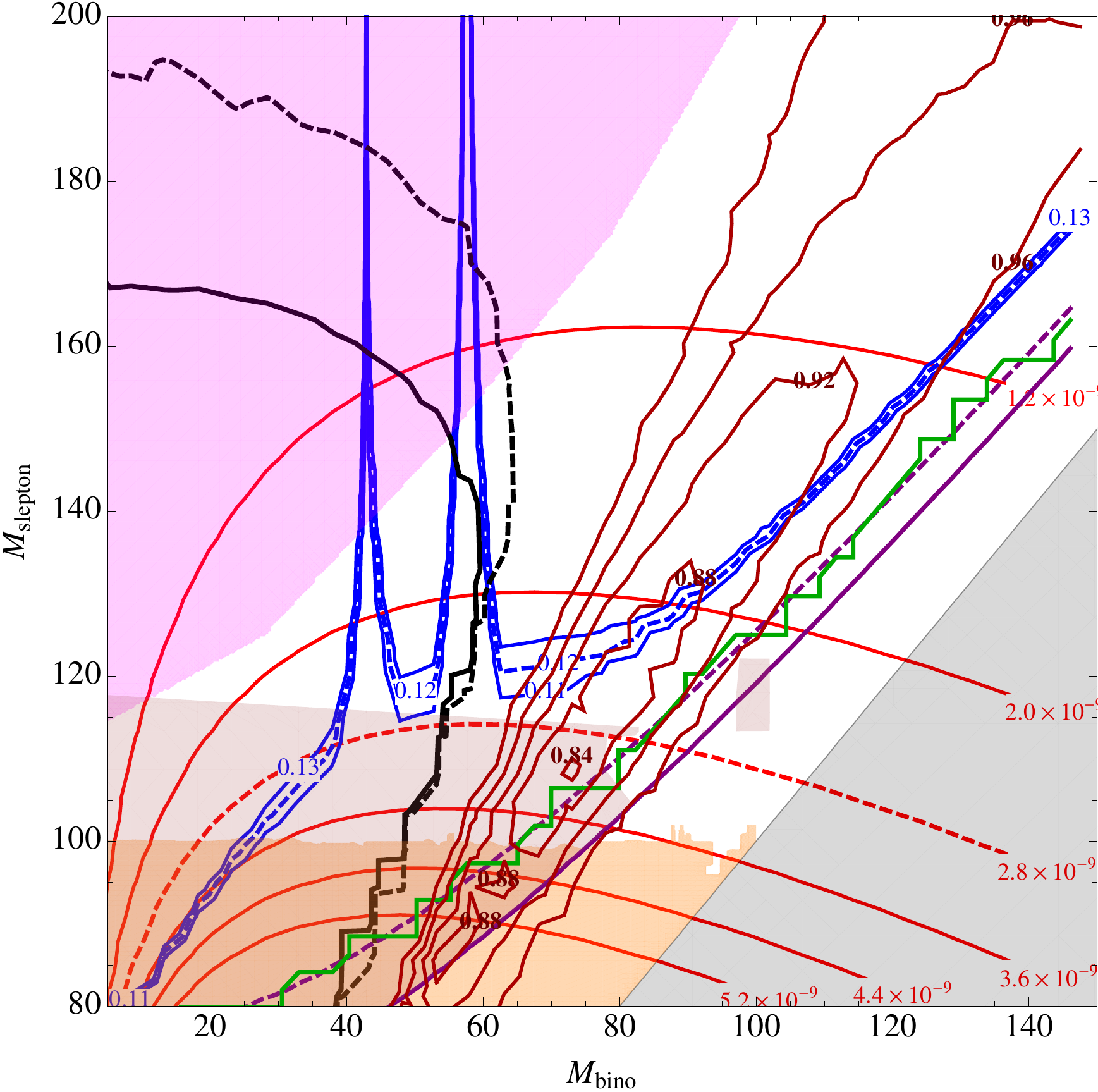}
&
\includegraphics[width=8.5cm]{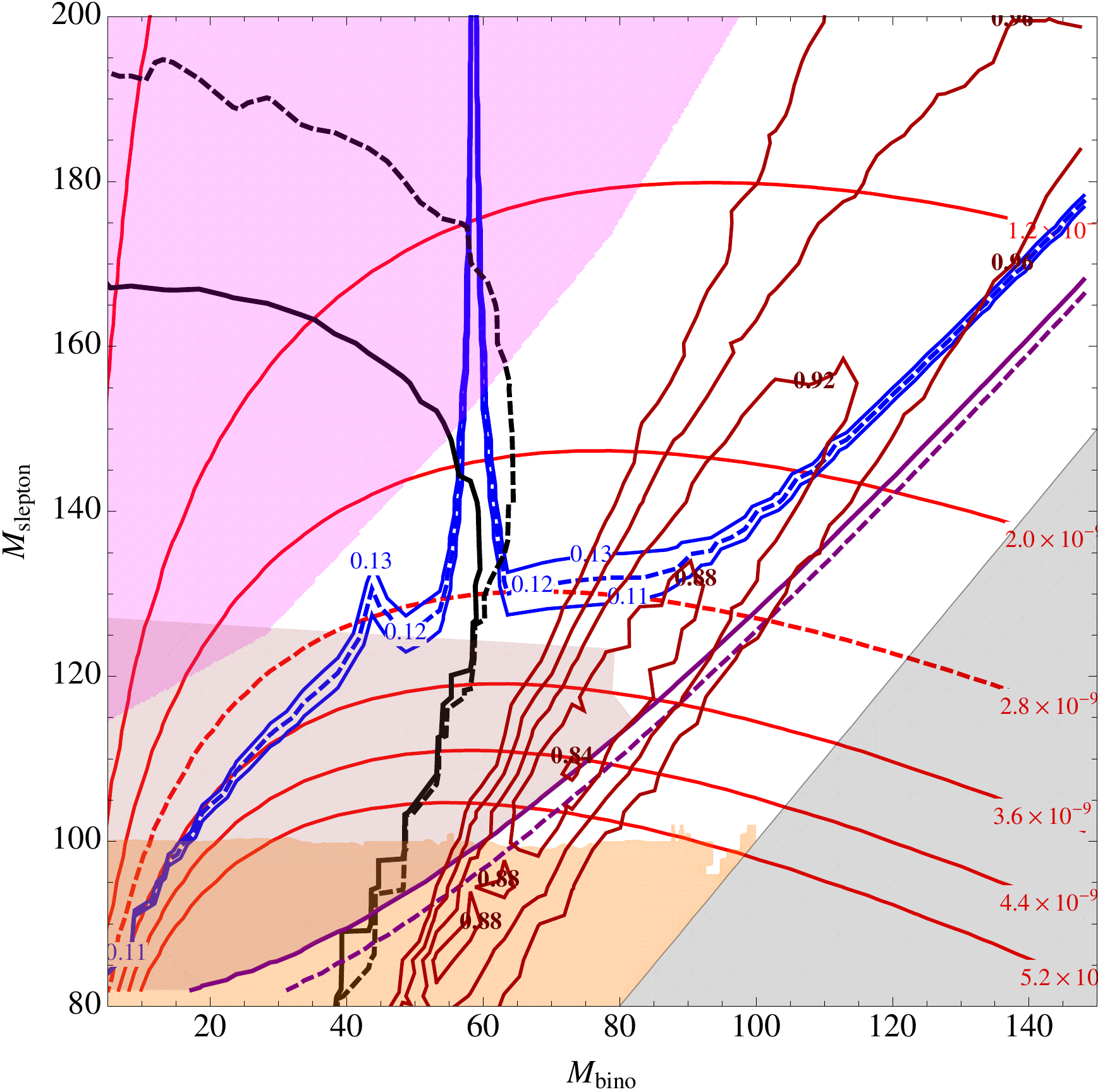}
\\
(c) $\tan \beta = 6, \mu = 400 \gev$ & (d) $\tan \beta = 6, \mu = 600 \gev$
\end{tabular} \vspace*{-9mm}
\end{center}
\caption{
Combination plot showing the overlap regions where our light slepton-bino scenario can account for both the 
$WW$-excess \cite{atlas7ww, cms7ww, cms8ww} (represented by dark red contours of $r_{\chi^2} < 1$),
the  DM relic density (blue dashed/solid contours: $\Omega_\mathrm{CDM} h^2 = 0.1196$ and $\pm 3 \times 0.0031 $) 
and $(g-2)_\mu$  (red dashed/solid contours: $\delta a_\mu=2.8 \times 10^{-9}$ and steps of $\pm0.8 \times 10^{-9}$, one $\sigma$ of the experimental measurement). 
The overlap region is centered around $M_\mathrm{bino} \approx 75 \gev$, $M_\mathrm{slepton} \approx 115 \gev$ for a range of $\mu$, $\tan \beta$. $A_\tau = 0, M_2 = 600 \gev$ in this plot,  and slepton soft mass universality with $m_{\tilde \ell_L} = m_{\tilde \ell_R}$ is assumed.
Grey (Orange) shaded regions are excluded by the LEP bound on $m_{\tilde \tau_1}$ ($m_{\tilde e, \tilde \mu}$)\cite{LEPlimits}. The magenta region is excluded by the CMS slepton search \cite{cmssleptons}, while black lines indicate our combined slepton bounds from the \ww cross section measurement, see Figs.~\ref{f.slepbounds} and \ref{f.chi2ratioww}. Regions below the solid (dashed) purple line have a stau (sneutrino) LSP.  Regions below the green line are excluded by the XENON100 direct detection bound \cite{xenon100} on the WIMP-nucleon cross section of $\sim 10^{-45} \mathrm{cm}^2$ for $M_{\mathrm{bino}} \approx 20 - 200 \gev$. 
}
\label{f.ponyplot}
\end{figure}

We have argued that the \ww excess, dark matter relic density and muon anomalous magnetic moment can all be explained by the same light slepton-bino scenario outlined at the beginning of this section. Light sleptons generate additional contributions to the \ww cross section measurement while ensuring a sufficiently large bino-annihilation cross section to ensure correct relic density. The region in the slepton-bino mass plane naturally gives the correct $\delta a_\mu$ contribution, and the small higgsino fraction of dark matter gives a direct detection cross section that is below current bounds but accessible to the next generation of experiments.  
As \fref{ponyplot} shows, all of this can be accommodated for moderately sized $\tan \beta \sim 5, \mu \sim 500 \gev$. Some \hgg enhancement can be accomodated by raising the stau soft masses and mixings and slightly heavier first and second generation sleptons.

The easiest way to experimentally test this scenario is to look for flavor correlations in the \ww cross section excess. Since sleptons cannot contribute to the $e \mu + \mathrm{MET}$ final state, comparing the \ww excess in that channel compared to the same-flavor channels should largely exclude (or support) this scenario with the full LHC run 1 dataset. This makes it different from the chargino scenario, because even if the same-sign dilepton signal predicted by \cite{Curtin:2012nn} is excluded, producing charginos via stops as in \cite{Rolbiecki:2013fia} could still account for the excess. In that case, more detailed study of possible discriminating kinematic variables is required.

%%%%%%%%%%%%%%%%%%%%%%%%%%%%%%%%%%%%%%%%%%%%%%%%%%%%%%%%%%%%%%%%%%%%%%
%%%%%%%%%%%%%%%%%%%%%%%%%%%%%%%%%%%%%%%%%%%%%%%%%%%%%%%%%%%%%%%%%%%%%%%
\section{Conclusion}
\label{s.conclusion} \setcounter{equation}{0} \setcounter{footnote}{0}
%%%%%%%%%%%%%%%%%%%%%%%%%%%%%%%%%%%%%%%%%%%%%%%%%%%%%%%%%%%%%%%%%%%%%%
%%%%%%%%%%%%%%%%%%%%%%%%%%%%%%%%%%%%%%%%%%%%%%%%%%%%%%%%%%%%%%%%%%%%%

By investigating just one SM standard candle, the \ww cross section, we have uncovered a wealth of possible information about new EW states at the LHC.  In the search for new EW states at the LHC prior to this study, a gap had consistently remained between LEP and the LHC for low mass EW states.  By examining the differential \ww cross section we have shown that this gap can be closed when investigating simplified models based on slepton production.  This work can be straightforwardly extended to other EW states such as charginos or non SUSY examples~\cite{workinprogress}.  In the slepton simplified model space we also discovered a region analogous to~\cite{Curtin:2012nn}, where new EW states can fit the \ww differential cross section data {\em better} than the SM alone.  Sleptons give significantly different predictions compared to the chargino~\cite{Curtin:2012nn} or stop~\cite{Rolbiecki:2013fia}  explanations of the \ww cross section anomaly, most notably that the excess in \ww should be flavor-diagonal. They can also account for the correct relic density of dark matter in the universe, provide a signal for future direct detection experiments, and explain the longstanding $(g-2)_\mu$ discrepancy.   Sleptons can also potentially explain some increase in the \hgg rate that may be slightly favored when combining both ATLAS and CMS results.  

Clearly any possible hints of new physics that SM standard candles shed light on should be investigated to the fullest.  However, regardless of whether or not this particular anomalous region remains with larger luminosity or higher energy runs, the importance of using SM standard candles is clear.  Current search strategies are typically based on looking in regions where the SM contributes a small number of events.  To investigate the actual EW scale, where the only new particle discovered by the LHC lurks, we must confront these regions by understanding the SM in greater detail.  Given that this prohibits the use of straightforward data-driven techniques it is important for  experimentalists to measure these regions in as much detail as possible, while theorists must continue to improve their calculations of SM processes.  Top physics has often been the hallmark of where to search for new physics, given that it couples so strongly to the Higgs.  However, EW gauge bosons also couple strongly to the source of EWSB, and can provide just as important of window into the physics associated with EWSB.  Both experimentalists and theorists need to explore the SM EW sector in exhaustive detail, otherwise we risk missing an important opportunity for discovering new physics, or understanding where it can and cannot exist.

\subsection*{Acknowledgements}

The work of D.C. was supported in part by the National Science Foundation under Grant PHY-0969739. The work of P.J. was supported in part by DOE under grant DE-FG02-97ER41022. The work of P.M.  and P.T.was supported in part by NSF CAREER Award NSF-PHY-1056833.
 
%%%%%%%%%%%%%%%%%%%%%%%%%%%%%%%%%%%%%%%%%%%%%%%%%%%%%%%%%%%%%%%%%%%%%%
%%%%%%%%%%%%%%%%%%%%%%%%%%%%%%%%%%%%%%%%%%%%%%%%%%%%%%%%%%%%%%%%%%%%%
%%%%%%%%%%%%%%%%%%%%%%%%%%%%%%%%%%%%%%%%%%%%%%%%%%%%%%%%%%%%%%%%%%%%%%
%%%%%%%%%%%%%%%%%%%%%%%%%%%%%%%%%%%%%%%%%%%%%%%%%%%%%%%%%%%%%%%%%%%%%

\end{document}